\newcommand*{\ATLASLATEXPATH}{}
\documentclass[UKenglish,texlive=2014,texmf,cernpreprint,PAPER]{\ATLASLATEXPATH atlasdoc}
\DeclareOldFontCommand{\rm}{\normalfont\rmfamily}{\mathrm}
\DeclareOldFontCommand{\sf}{\normalfont\sffamily}{\mathsf}
\DeclareOldFontCommand{\tt}{\normalfont\ttfamily}{\mathtt}
\DeclareOldFontCommand{\bf}{\normalfont\bfseries}{\mathbf}
\DeclareOldFontCommand{\it}{\normalfont\itshape}{\mathit}
\DeclareOldFontCommand{\sl}{\normalfont\slshape}{\@nomath\sl}
\DeclareOldFontCommand{\sc}{\normalfont\scshape}{\@nomath\sc}

\usepackage{\ATLASLATEXPATH atlaspackage}
\usepackage{\ATLASLATEXPATH atlasbiblatex}

\usepackage{\ATLASLATEXPATH atlascontribute}

\usepackage{\ATLASLATEXPATH atlasphysics}

\usepackage{nicefrac}
\graphicspath{{logos/}{figures/}}

\newcommand{\pPb}{\mbox{$p$+Pb}}
\newcommand{\sqrtsnn}{\sqrt{s_{\mathrm{NN}}}}
\newcommand{\invnb}{\mathrm{nb}^{-1}}

\newcommand{\lr}[1]{\left\langle #1\right\rangle}

\newcommand{\nchrec}{N_{\mathrm{ch}}^{\mathrm{rec}}}
\newcommand{\nch}{N_{\mathrm{ch}}}

\AtlasTitle{Measurement of forward-backward multiplicity correlations in lead-lead, proton-lead and proton-proton collisions with the ATLAS detector}

\author{The ATLAS Collaboration}

\PreprintIdNumber{CERN-PH-2016-124}

\AtlasDate{\today}

\AtlasJournalRef{Phys. Rev. C 95 (2017) 064914}
\AtlasDOI{10.1103/PhysRevC.95.064914}
\arXivId{1606.08170}

\AtlasAbstract{Two-particle pseudorapidity correlations are measured in $\sqrt{s_{\rm{NN}}}$~=~2.76 $\TeV$ Pb+Pb,\\               $\sqrt{s_{\rm{NN}}}$~=~5.02~$\TeV$ $p$+Pb, and $\sqrt{s}$~=~13 $\TeV$ $pp$ collisions at the LHC, with total integrated luminosities of approximately 7 $\mu\mathrm{b}^{-1}$, 28 $\mathrm{nb}^{-1}$, and 65 $\mathrm{nb}^{-1}$, respectively. The correlation function $C_{\rm N}(\eta_1,\eta_2)$ is measured as a function of event multiplicity using charged particles in the pseudorapidity range $|\eta|<2.4$. The correlation function contains a significant short-range component, which is estimated and subtracted. After removal of the short-range component, the shape of the correlation function is described approximately by $1+\langle{a_1^2}\rangle^{\nicefrac{1}{2}} \eta_1\eta_2$ in all collision systems over the full multiplicity range. The values of $\langle{a_1^2}\rangle^{\nicefrac{1}{2}}$ are consistent between the opposite-charge pairs and same-charge pairs, and for the three collision systems at similar multiplicity. The values of $\langle{a_1^2}\rangle^{\nicefrac{1}{2}}$ and the magnitude of the short-range component both follow a power-law dependence on the event multiplicity. The short-range component in $p$+Pb collisions, after symmetrizing the proton and lead directions, is found to be smaller at a given $\eta$ than in $pp$ collisions with comparable multiplicity.}

\hypersetup{pdftitle={ATLAS document},pdfauthor={The ATLAS Collaboration}}

\begin{document}

\maketitle

\tableofcontents


\section{Introduction}
\label{intro}

Heavy-ion collisions at RHIC and the LHC create hot, dense matter whose space-time evolution can be well described by relativistic viscous hydrodynamics~\cite{Gale:2013da,Heinz:2013th}. Owing to strong event-by-event (EbyE) density fluctuations in the initial state, the space-time evolution of the produced matter in the final state also fluctuates event to event. These fluctuations may lead to correlations of particle multiplicity in momentum space in the transverse and longitudinal directions with respect to the collision axis. Studies of the multiplicity correlation in the transverse plane have revealed strong harmonic modulation of the particle densities in the azimuthal angle, commonly referred to as harmonic flow. The measurements of harmonic flow coefficients $v_n$~\cite{Adare:2011tg,ALICE:2011ab,Aad:2012bu,Chatrchyan:2013kba} and their EbyE fluctuations~\cite{Aad:2013xma,Aad:2014fla,Aad:2015lwa,Adam:2015eta} have placed important constraints on the properties of the medium and transverse energy density fluctuations in the initial state.

Two-particle correlations in the transverse plane have also been studied in high-multiplicity $\pp$~\cite{Khachatryan:2010gv,Aad:2015gqa,Khachatryan:2015lva} and $\pPb$~\cite{CMS:2012qk,Abelev:2012ola,Aad:2012gla,Chatrchyan:2013nka,Aad:2014lta} collisions, and these studies have revealed features that bear considerable similarity to those observed in heavy-ion collisions. These findings have generated many theoretical interpretations~\cite{Dusling:2015gta}, and much discussion as to whether the mechanisms that result in the observed correlations are or are not fundamentally the same in the different collision systems. 

This paper reports measurements of multiplicity correlations in the longitudinal direction in $\pp$, $\pPb$, and Pb+Pb collisions, which are sensitive to the early-time density fluctuations in pseudorapidity ($\eta$)~\cite{Gale:2013da,Heinz:2013th}. These density fluctuations generate long-range correlations (LRC) at the early stages of the collision, well before the onset of any collective behavior, and appear as correlations of the multiplicity densities of produced particles separated in $\eta$. For example, the EbyE differences between the partonic flux in the target and the projectile may lead to a long-range asymmetry of the produced system~\cite{Bialas:2011bz,Bzdak:2012tp,Jia:2014ysa}, which manifests itself as a correlation between the multiplicity densities of final-state particles with large $\eta$ separation. 

Longitudinal multiplicity correlations can also be generated during the space-time evolution in the final state as resonance decays, single-jet fragmentation, and Bose-Einstein correlations. These latter correlations are typically localized over a smaller range of $\eta$, and are commonly referred to as short-range correlations (SRC). On the other hand, di-jet fragmentation may contribute to the LRC if the $\eta$ separation between the two jets is large.

Many previous studies are based on forward-backward (FB) correlations of particle multiplicity in two $\eta$ ranges symmetric around the center-of-mass of the collision systems, including $e^+e^-$~\cite{Braunschweig:1989bp}, $\pp$~\cite{Uhlig:1977dc,Ansorge:1988fg,ATLAS:2012as,Adam:2015mya}, and A+A~\cite{Back:2006id,Abelev:2009ag} collisions where a significant anti-correlation between forward and backward multiplicities has been identified. Recently, the study of multiplicity correlations has been generalized by decomposing the correlation function into orthogonal Legendre polynomial functions, or more generally into principal components, each representing a unique component of the measured FB correlation~\cite{Bzdak:2012tp,Bhalerao:2014mua}. 

Particle production in $\pp$ collisions is usually described by QCD-inspired models, such as PYTHIA~\cite{Sjostrand:2007gs} and EPOS~\cite{Pierog:2013ria}, implemented in Monte Carlo (MC) event generators with free parameters that are tuned to describe experimental measurements. Previous studies show that these models can generally describe the $\eta$ and $\pT$ dependence of the inclusive charged-particle production~\cite{Aad:2016mok,Aaboud:2016itf}, as well as the underlying event accompanying various hard-scattering processes~\cite{Aad:2014hia,Aad:2014jgf}.  In many such models, events with large charged-particle multiplicity are produced through multiple parton-parton interactions (MPI), which naturally serve as sources for the FB multiplicity asymmetry describe above. Therefore, a detailed measurement of pseudorapidity correlation in $\pp$ collisions also provides new constraints on the longitudinal dynamics of MPI processes in these models.

The two-particle correlation function in pseudorapidity is defined as~\cite{Berger:1974vn,jjia}:
\begin{eqnarray}
\label{eq:1}
C(\eta_1,\eta_2) = \frac{\left\langle N(\eta_1) N(\eta_2)\right\rangle}{\left\langle N(\eta_1)\right\rangle\left\langle N(\eta_2)\right\rangle}\equiv \left\langle \rho(\eta_1)\rho(\eta_2)\right\rangle\;,\;\;\; \rho(\eta)\equiv \frac{ N(\eta)}{\left\langle N(\eta)\right\rangle}\;,
\end{eqnarray}
where $N(\eta)$ is the multiplicity density distribution in a single event and $\lr{N(\eta)}$ is the average distribution for a given event-multiplicity class. The correlation function is directly related to a single-particle quantity $\rho(\eta)$, which characterizes the fluctuation of multiplicity in a single event relative to the average shape of the event class.

Following Refs.~\cite{Bzdak:2012tp,jjia},  $\rho(\eta)$ in the interval [$-Y$,$Y$] is written in terms of Legendre polynomials:
\begin{eqnarray}\label{eq:2}
\rho(\eta) &\propto& 1+\sum_na_n\; T_n(\eta)\;, T_n(\eta) \equiv \sqrt{\frac{2n+1}{3}} Y\; P_n\left(\frac{\eta}{Y}\right)\;,
\end{eqnarray}
and the scale factor in Eq.~\eqref{eq:2} is chosen such that $T_1(\eta)=\eta$.~\footnote{The $T_n(\eta)$ also satisfy: $\int_{-Y}^{Y}T_n(\eta)d\eta=0$ for $n\geq 1$, and $\int_{-Y}^{Y}T_n(\eta)T_m(\eta)d\eta=\left(\frac{2Y^2}{3}\delta_{nm}\right)$. From the definition of $\rho(\eta)$ in Eq.~\eqref{eq:1}, it follows that $\lr{\sum_{n=0}^{\infty} a_nT_n(\eta)}=0$.} 

Using Eqs.~\eqref{eq:1} and \eqref{eq:2}, the correlation function $C$ can be expressed in terms of the $T_n$, which involve terms in $\lr{a_0a_0}$, $\lr{a_0a_n}$, and $\lr{a_na_m}$, with $n,m\geq1$. Terms involving $a_0$ reflect multiplicity fluctuations in the given event class, while the dynamical fluctuations between particles at different pseudorapidities in events of fixed multiplicity are captured by the terms in $\lr{a_na_m}$, $n,m\geq1$. It is the study of these dynamical fluctuations that is the goal of this analysis.

As discussed in more detail in Ref.~\cite{jjia}, the terms involving $\lr{a_0a_n}$ can be removed, provided all  deviations from 1 are small, by defining:
\begin{equation} 
\label{eq:3}
C_{\rm N}(\eta_1,\eta_2) = \frac{C(\eta_1,\eta_2)}{C_{p}(\eta_1)C_{p}(\eta_2)}\;,
\end{equation}
where 
\begin{equation} 
\label{eq:4}
C_{p}(\eta_1) = \frac{\int_{-Y}^{Y} C(\eta_1,\eta_2) \,d\eta_2}{2Y}\;,
\end{equation}
with a similar expression for $C_{p}(\eta_2)$. The quantities $C_{p}(\eta_1)$ and $C_{p}(\eta_2)$ are referred to as the single-particle modes. The $\lr{a_0a_0}$ term can be removed by renormalizing average value in the $\eta_1$, $\eta_2$ phase space to be 1. The final result is:
\begin{eqnarray}\label{eq:5}
C_{\rm N}(\eta_1,\eta_2)= 1+\sum_{n,m=1}^{\infty} a_{n,m} \frac{T_n(\eta_1)T_m(\eta_2)+T_n(\eta_2)T_m(\eta_1)}{2},\; \mbox{and}\;a_{n,m}\equiv\lr{a_na_m}\;.  
\end{eqnarray}
The two-particle Legendre coefficients can be calculated directly from the measured correlation function:
\begin{eqnarray}\label{eq:6}
a_{n,m} = \left(\frac{3}{2Y^3}\right)^2\int_{-Y}^{Y} C_{\rm N}(\eta_1,\eta_2) \frac{T_n(\eta_1)T_m(\eta_2)+T_n(\eta_2)T_m(\eta_1)}{2} \,d\eta_1\,d\eta_2\;.
\end{eqnarray}
The two-particle correlation method measures, in effect, the root-mean-square (RMS) values of the EbyE $a_n$, $\lr{a_n^2}^{\nicefrac{1}{2}}$, or the cross correlation between $a_n$ and $a_m$, $\lr{a_na_m}$. The correlation functions satisfy the symmetry condition $C(\eta_1,\eta_2) =C(\eta_2,\eta_1)$ and $C_{\rm N}(\eta_1,\eta_2) =C_{\rm N}(\eta_2,\eta_1)$.

This paper presents a measurement of the two-dimensional (2-D) correlation function $C_{\rm N}(\eta_1,\eta_2)$ over the pseudorapidity range of $|\eta|<2.4$ in $\sqrtsnn$~=~2.76 $\TeV$ Pb+Pb, $\sqrtsnn$~=~5.02~$\TeV$ $\pPb$, and $\sqrt{s}$~=~13~$\TeV$ $\pp$ collisions, using the ATLAS detector.\footnote{ATLAS uses a right-handed coordinate system with its origin at the nominal interaction point (IP) in the center of the detector and the $z$-axis along the beam pipe. The $x$-axis points from the IP to the center of the LHC ring, and the $y$-axis points upward. Cylindrical coordinates $(r,\phi)$ are used in the transverse plane, $\phi$ being the azimuthal angle around the beam pipe. The pseudorapidity is defined in terms of the polar angle $\theta$ as $\eta=-\ln\tan(\theta/2)$.} The analysis is performed using events for which the total number of reconstructed charged particles, $\nchrec$, with $|\eta|<2.5$ and transverse momentum $\pT>0.4$~$\GeV$, is in the range $10\leq \nchrec < 300$. Both the Pb+Pb and $\pPb$ data cover this range of $\nchrec$, but for $\pp$ the range extends only to approximately 160. The measured $C_{\rm N}(\eta_1,\eta_2)$ is separated into a short-range component $\delta_{\rm {SRC}}(\eta_1,\eta_2)$ and $C_{\rm N}^{\mathrm{sub}}(\eta_1,\eta_2)$, which contains the long-range component. The nature of the FB fluctuation in each collision system is studied by projections as well as Legendre coefficients $\lr{a_na_m}$ of $C_{\rm N}^{\mathrm{sub}}(\eta_1,\eta_2)$. The magnitudes of the FB fluctuations are compared for the three systems at similar event multiplicity. A comparison is also made between the $\pp$ data and QCD-inspired models.

\section{ATLAS detector and trigger}
The ATLAS detector~\cite{Aad:2008zzm} provides nearly full solid-angle coverage of the collision point with tracking detectors, calorimeters, and muon chambers, and is well suited for measurement of two-particle correlations over a large pseudorapidity range. The measurements were performed using the inner detector (ID), minimum-bias trigger scintillators (MBTS), the forward calorimeter (FCal), and the zero-degree calorimeters (ZDC). The ID detects charged particles within $|\eta| < 2.5$ using a combination of silicon pixel detectors, silicon microstrip detectors~(SCT), and a straw-tube transition radiation tracker (TRT), all immersed in a 2 T axial magnetic field~\cite{Aad:ID}. An additional pixel layer, the ``Insertable B Layer'' (IBL)~\cite{atlas:1,atlas:2} installed between Run 1 and Run 2 (2013--2015), is used in the 13~$\TeV$ $\pp$ measurements. The MBTS system detects charged particles over $2.1\lesssim|\eta|\lesssim3.9$ using two hodoscopes of counters positioned at $z = \pm$~3.6 m. The FCal consists of three sampling layers, longitudinal in shower depth, and covers $3.2<|\eta|< 4.9$. The ZDC, available in the Pb+Pb and $\pPb$ runs, are positioned at $\pm$140~m from the collision point, detecting neutrons and photons with $|\eta|>8.3$. 

This analysis uses approximately 7~$\mu\mathrm{b}^{-1}$ of Pb+Pb data, 28~$\invnb$ of $\pPb$ data, and 65 $\invnb$ of $\pp$ data taken by the ATLAS experiment at the LHC. The Pb+Pb data were collected in 2010 at a nucleon-nucleon center-of-mass energy $\sqrtsnn=2.76$ $\TeV$. The $\pPb$ data were collected in 2013, when the LHC was configured with a 4~$\TeV$ proton beam and a 1.57~$\TeV$ per-nucleon Pb beam that together produced collisions at $\sqrtsnn = 5.02$~$\TeV$. The higher energy of the proton beam results in a rapidity shift of $0.47$ of the nucleon-nucleon center-of-mass frame towards the proton beam direction relative to the laboratory rest frame. The $\pp$ data were collected during a low-luminosity operation of the LHC in June and August of 2015 at collision energy $\sqrt{s} = 13$~$\TeV$. 

The ATLAS trigger system~\cite{Aad:2012xs} consists of a Level-1 (L1) trigger implemented using a combination of dedicated electronics and programmable logic, and a high-level trigger (HLT) implemented in processors. The HLT reconstructs charged-particle tracks using methods similar to those applied in the offline analysis, allowing high-multiplicity track (HMT) triggers that select on the number of tracks having $\pT >$~0.4~$\GeV$ associated with a vertex with largest number of associated tracks (primary vertex). The Pb+Pb data used in the analysis are collected by a minimum-bias trigger, while the $\pp$ and $\pPb$ data are collected by a minimum-bias trigger and HMT triggers. 

The Pb+Pb trigger requires signals in two ZDCs or either of the two MBTS counters. The ZDC trigger thresholds on each side are set below the peak corresponding to a single neutron. A timing requirement based on signals from each side of the MBTS is imposed to remove beam backgrounds. The minimum-bias trigger for $\pPb$ is similar, except that only the ZDC on the Pb-fragmentation side is used. For $\pp$ collisions, the minimum-bias trigger requires only one or more signals in the MBTS.

Two distinct HMT triggers are used for the 13~$\TeV$ $\pp$ analysis. The first trigger selected events at L1 that have a signal in at least one counter on each side of the MBTS, and at the HLT have at least 900 SCT hits and 60 tracks associated with a primary vertex. The second trigger selects events with a total transverse energy of more than 10~$\GeV$ at L1 and at least 1400 SCT hits and 90 tracks associated to a primary vertex at HLT. For the $\pPb$ data, the HMT triggers were formed from a combination of L1 triggers that applied different thresholds for total transverse energy measured over $3.2< |\eta| < 4.9$ in the FCal and HLT triggers that placed minimum requirements on the number of reconstructed tracks. Details of the minimum-bias and HMT triggers can be found in Refs.~\cite{Aad:2016mok,Aad:2015gqa} and Refs.~\cite{Aad:2014lta,atlas:7} for the $\pp$ and $\pPb$ collisions, respectively.

\section{Data analysis}
\label{sec:4}
\subsection{Event and track selection}
\label{sec:sel}
The offline event selection for the $\pPb$ and $\pp$ data requires at least one reconstructed vertex with its $z$ position satisfying $|\zvtx|< 100$~mm. The mean collision rate per crossing $\mu$ is around 0.03 for $\pPb$ data, between $0.002$ and $0.04$ for the June 2015 $\pp$ data, and between 0.05 and 0.6 for the August 2015 $\pp$ data. Events containing multiple collisions (pileup) are suppressed by rejecting events with more than one good reconstructed vertex, and results are found to be consistent between the June and August datasets. For the $\pPb$ events, a time difference of $|\Delta t| < 10$ ns is also required between signals in the MBTS counters on either side of the interaction point to suppress noncollision backgrounds. 

 The offline event selection for the Pb+Pb data requires a reconstructed vertex with its $z$ position satisfying $|\zvtx|< 100$~mm. The selection also requires a time difference $|\Delta t| < 3$ ns between signals in the MBTS trigger counters on either side of the interaction point to suppress noncollision backgrounds. A coincidence between the ZDC signals at forward and backward pseudorapidity is required to reject a variety of background processes, while maintaining more than 98\% efficiency for inelastic processes.

Charged-particle tracks and primary vertices are reconstructed in the ID using algorithms whose implementation was optimized for better performance between LHC Runs 1 and 2. In order to compare directly the $\pPb$ and Pb+Pb systems using event selections based on the multiplicity of the collisions, a subset of data from peripheral Pb+Pb collisions, collected during the 2010 LHC heavy-ion run with a minimum-bias trigger, was reanalyzed using the same track reconstruction algorithm as that used for $\pPb$ collisions. For the $p$+Pb and Pb+Pb analyses, tracks are required to have a $\pT$-dependent minimum number of hits in the SCT, and the transverse ($d_0$) and longitudinal ($z_0$ $\sin\theta$) impact parameters of the track relative to the vertex are required to be less than 1.5 mm. A description of the 2010 Pb+Pb data and 2013 $\pPb$ data can be found in Ref.~\cite{Aad:2012bu} and Ref.~\cite{atlas:5}, respectively.

For the 13~$\TeV$ $\pp$ analysis, the track selection criteria were modified slightly to profit from the presence of the IBL in Run 2. Furthermore, the requirements of $|d_0^{\rm{z}}|<1.5$ mm and $|z_0\sin\theta|<1.5$~mm are applied, where $d_0^z$ is the transverse impact parameter of the track relative to the average beam position. These selection criteria are the same as those in Refs.~\cite{Aad:2016mok,Aad:2015gqa}.

\begin{figure}[!h]
\begin{center}
\includegraphics[width=0.6\linewidth]{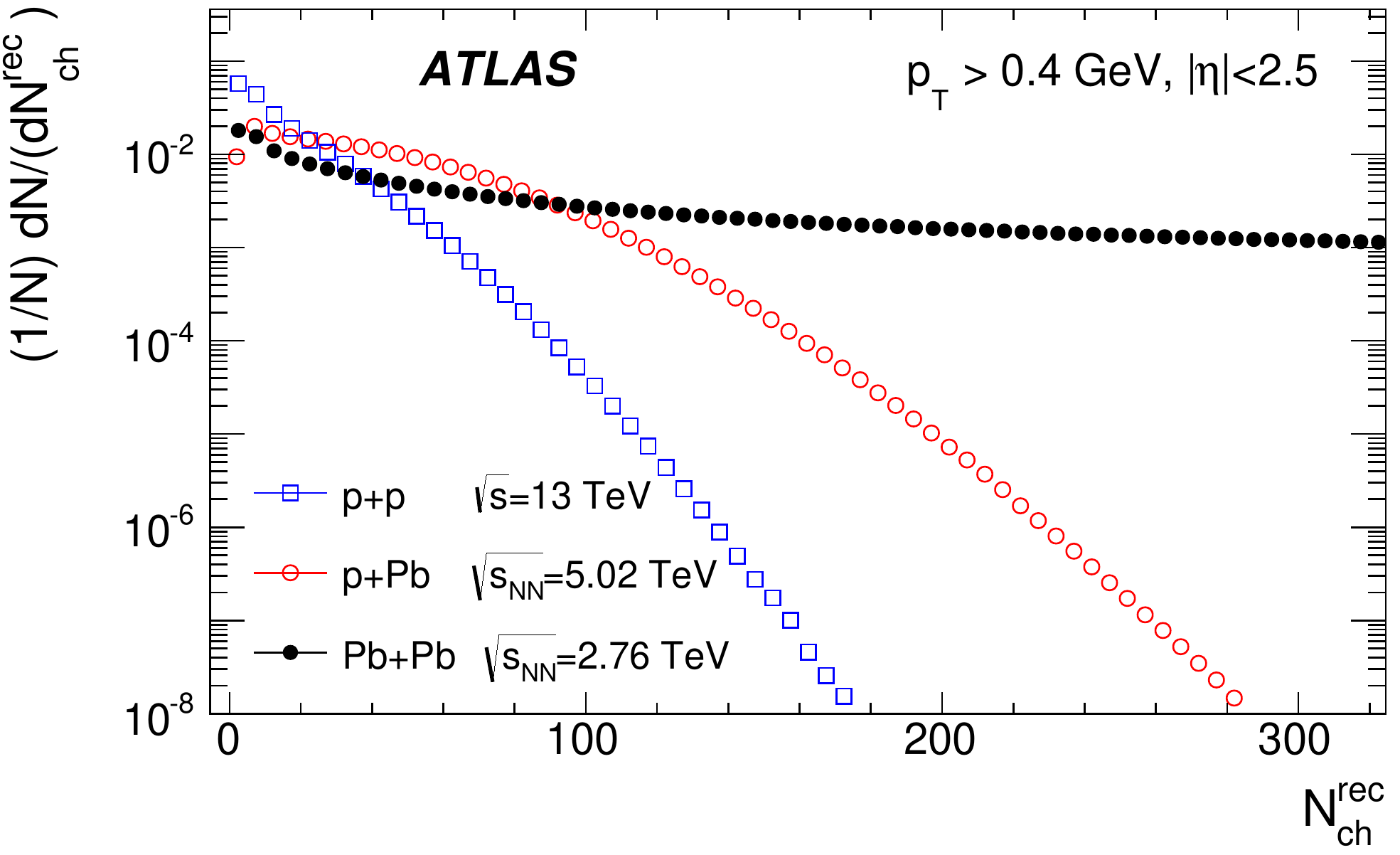}
\end{center}
\caption{\label{fig:0} The normalized distributions of the number of reconstructed tracks, $\nchrec$, with $\pT > 0.4$~$\GeV$ and $|\eta|<2.5$ in the three collision systems. The $N_{\rm{evts}}$ refers to the number of collisions for a given $\nchrec$.} 
\end{figure}

In this analysis, the correlation functions are constructed using tracks passing the above selection requirements and which have $\pT >$~0.2~$\GeV$ and $|\eta|<2.4$. However, slightly different kinematic requirements, $\pT >$ 0.4~$\GeV$ and $|\eta|<2.5$, are used to count the number of reconstructed charged particles in the event, denoted by $\nchrec$, to be consistent with the requirements used in the HLT. Figure~\ref{fig:0} compares the normalized $\nchrec$ distributions of events in the three colliding systems. The distribution decreases slowly in the Pb+Pb system, but decreases much faster in the $\pPb$ and $\pp$ systems. A major goal of the analysis is to compare the correlation function from the three collisions systems at similar $\nchrec$ values, which can reveal whether the FB multiplicity fluctuation is controlled by the collision geometry or the overall activity of the event.

The efficiency of the track reconstruction and track selection requirements, $\epsilon(\eta,\pT)$, is evaluated using simulated $\pPb$ or Pb+Pb events produced with the HIJING event generator~\cite{Gyulassy:1994ew} or simulated $\pp$ events from the PYTHIA 8~\cite{Sjostrand:2007gs} event generator using parameter settings according to the so-called A2 tune~\cite{atlas:6}. The MC sample for Pb+Pb events in the multiplicity region of interest was very small, therefore the reconstruction efficiency for Pb+Pb was taken from the larger $\pPb$ sample. The $\pPb$ efficiency was found to be consistent with the efficiency from the Pb+Pb MC simulation, but of much higher precision. The response of the detector to these MC events is simulated using GEANT4~\cite{Agostinelli:2002hh,Aad:2010ah} and the resulting events are reconstructed with the same algorithms that are applied to the data. The efficiencies for the three datasets are similar for events with similar multiplicity. Small differences are due to changes in the detector conditions in Run 1 and changes in the reconstruction algorithm between Runs 1 and 2. In the simulated events, the efficiency reduces the measured charged-particle multiplicity relative to the event generator multiplicity for primary charged particles.~\footnote{For Pb+Pb and $\pPb$ simulation, the event generator multiplicity includes charged particles that originate directly from the collision or result from decays of particles with $c\tau<$ 10 mm. The definition for primary charged particles is somewhat stronger in the $\pp$ simulation~\cite{Aad:2016mok}.} The reduction factors for $\nchrec$ and the associated efficiency uncertainties are $b=1.29$ $\pm$ 0.05, 1.29 $\pm$ 0.05, and 1.18~$\pm$~0.05 for Pb+Pb, $\pPb$, and $\pp$ collisions, respectively. The values of these reduction factors are found to be independent of multiplicity over the $\nchrec$ range used in this analysis, $10\leq\nchrec<300$. Therefore, these factors are used to multiply $\nchrec$ to obtain the efficiency-corrected average number of charged particles with $\pT > 0.4$~$\GeV$ and $|\eta|<2.5$, $\nch=b\nchrec$. The quantity $\nch$ is used when presenting the multiplicity dependence of the SRC and the LRC.

\subsection{Two-particle correlations}
\label{sec:corr}
The two-particle correlation function defined in Eq.~\eqref{eq:1} is calculated as the ratio of distributions for same-event pairs $S(\eta_1,\eta_2)\propto\lr{N(\eta_1)N(\eta_2)}$, and mixed-event pairs $B(\eta_1,\eta_2)\propto \lr{N(\eta_1)}\lr{N(\eta_2)}$~\cite{Aad:2012bu}:
\begin{equation} 
\label{eq:b0}
C(\eta_1,\eta_2) = \frac{S(\eta_1,\eta_2)}{B(\eta_1,\eta_2)}\;.
\end{equation}
The mixed-event pair distribution is constructed by combining tracks from one event with those from another event with similar $\nchrec$ (matched within two tracks) and $z_{\mathrm{vtx}}$ (matched within 2.5~mm). The events are also required to be close to each other in time to account for possible time-dependent variation of the detector conditions. The mixed-event distribution should account properly for detector inefficiencies and non-uniformity, but does not contain physical correlations. The normalization of $C(\eta_1,\eta_2)$ is chosen such that its average value in the $(\eta_1, \eta_2)$ plane is one. The correlation function satisfies the symmetry $C(\eta_1,\eta_2)=C(\eta_2,\eta_1)$ and, for a symmetric collision system, $C(\eta_1,\eta_2)~=~C(-\eta_1,-\eta_2)$. Therefore, for $\pp$ and Pb+Pb collisions, all pairs are entered into one quadrant of the $(\eta_1, \eta_2)$ space defined by $\eta_- \equiv \eta_1-\eta_2>0$ and $\eta_+~\equiv~\eta_1~+~\eta_2~>~0$ and then reflected to the other quadrants. For $\pPb$ collisions, all pairs are entered into one half of the $(\eta_1, \eta_2)$ space defined by  $\eta_1-\eta_2>0$ and then reflected to the other half. To correct $S(\eta_1,\eta_2)$ and $B(\eta_1,\eta_2)$ for the individual inefficiencies of particles in the pair, the pairs are weighted by the inverse product of their tracking efficiencies $1/(\epsilon_1\epsilon_2)$. Remaining detector distortions not accounted for by the reconstruction efficiency largely cancel in the same-event to mixed-event ratio.

In a separate analysis, the correlation functions in $\pPb$ collisions are also symmetrized in the same way as for Pb+Pb and $\pp$ collisions such that $C(\eta_1,\eta_2)=C(-\eta_1,-\eta_2)$, and they are compared with correlation functions obtained for symmetric collision systems. This symmetrized $\pPb$ correlation function is used only at the end of Sec.~\ref{sec:result}, in relation to Fig.~\ref{fig:13}. In all other cases the $\pPb$ correlation function is unsymmetrized.

\subsection{Outline of the procedure for separating SRC and LRC}
\label{sec:33}
As explained in the introduction, the aim of this analysis is to measure and parametrize the long-range correlation, which requires the separation and subtraction of the short-range component. The separation of SRC and LRC is quite involved and so is briefly summarized here, with details left to the relevant later sections.

The core of the separation method is to exploit the difference between the correlations for opposite-charge and same-charge pairs, $C^{+-}(\eta_1,\eta_2)$ and $C^{\pm\pm}(\eta_1,\eta_2)$, respectively. The SRC component centered around $\eta_-(\equiv\eta_1-\eta_2)\sim0$ is found to be much stronger for opposite-charge pairs, primarily due to local charge conservation, while the LRC and single-particle modes are expected to be independent of the charge combination. With this assumption, the ratio:
\begin{eqnarray}
\label{eq:c0}
R(\eta_1,\eta_2)= C^{+-}(\eta_1,\eta_2)/C^{\pm\pm}(\eta_1,\eta_2)
\end{eqnarray}
is given approximately by:
\begin{eqnarray}
\label{eq:c1}
R(\eta_1,\eta_2) \approx 1 + \delta_{\mathrm{SRC}}^{+-}(\eta_1,\eta_2) -\delta_{\mathrm{SRC}}^{\pm\pm}(\eta_1,\eta_2)
\end{eqnarray}
This analysis assumes further that the dependence of $\delta_{\mathrm{SRC}}$ on $\eta_-$ and $\eta_+(\equiv\eta_1+\eta_2)$ factorizes, and that the dependence on $\eta_+$ is independent of the charge combination: $\delta_{\mathrm{SRC}}^{+-}= f(\eta_+)g^{+-}(\eta_-)$, $\delta_{\mathrm{SRC}}^{\pm\pm}= f(\eta_+)g^{\pm\pm}(\eta_-)$, where $g^{+-}(\eta_-)$ and $g^{\pm\pm}(\eta_-)$ are allowed to differ in both shape and magnitude. With these assumptions~\footnote{The validity of the various assumptions is confirmed in the data from the extracted $\delta_{\mathrm{SRC}}^{+-}(\eta_+,\eta_-)$ and $\delta_{\mathrm{SRC}}^{\pm\pm}(\eta_+,\eta_-)$ after applying the separation procedure.}, $f(\eta_+)$ can be determined from $R$ by suitable integration over $\eta_-$, as described in Sec.~\ref{sec:34}.

To complete the determination of $\delta_{\mathrm{SRC}}^{\pm\pm}$, the quantity $g^{\pm\pm}$ is determined and parameterized from suitable projections of $C_{\rm N}^{\pm\pm}(\eta_+,\eta_-)$ in the $\eta_-$ direction, as described in Sec.~\ref{sec:sub}. The use of $C_{\rm N}^{\pm\pm}$ rather than $C^{\pm\pm}$ is because the former does not contain the single-particle modes. The procedure to obtain a correlation function with the SRC subtracted is also described in Sec.~\ref{sec:sub}. With $\delta_{\mathrm{SRC}}^{\pm\pm}$ determined, $\delta_{\mathrm{SRC}}^{+-}$ is obtained directly from Eq.~\eqref{eq:c1}. The $\delta_{\mathrm{SRC}}^{\pm\pm}$ and $\delta_{\mathrm{SRC}}^{+-}$ are then averaged to obtain the SRC for all charge combinations, $\delta_{\mathrm{SRC}}$.

\begin{figure}[!t]
\begin{center}
\includegraphics[width=1\linewidth]{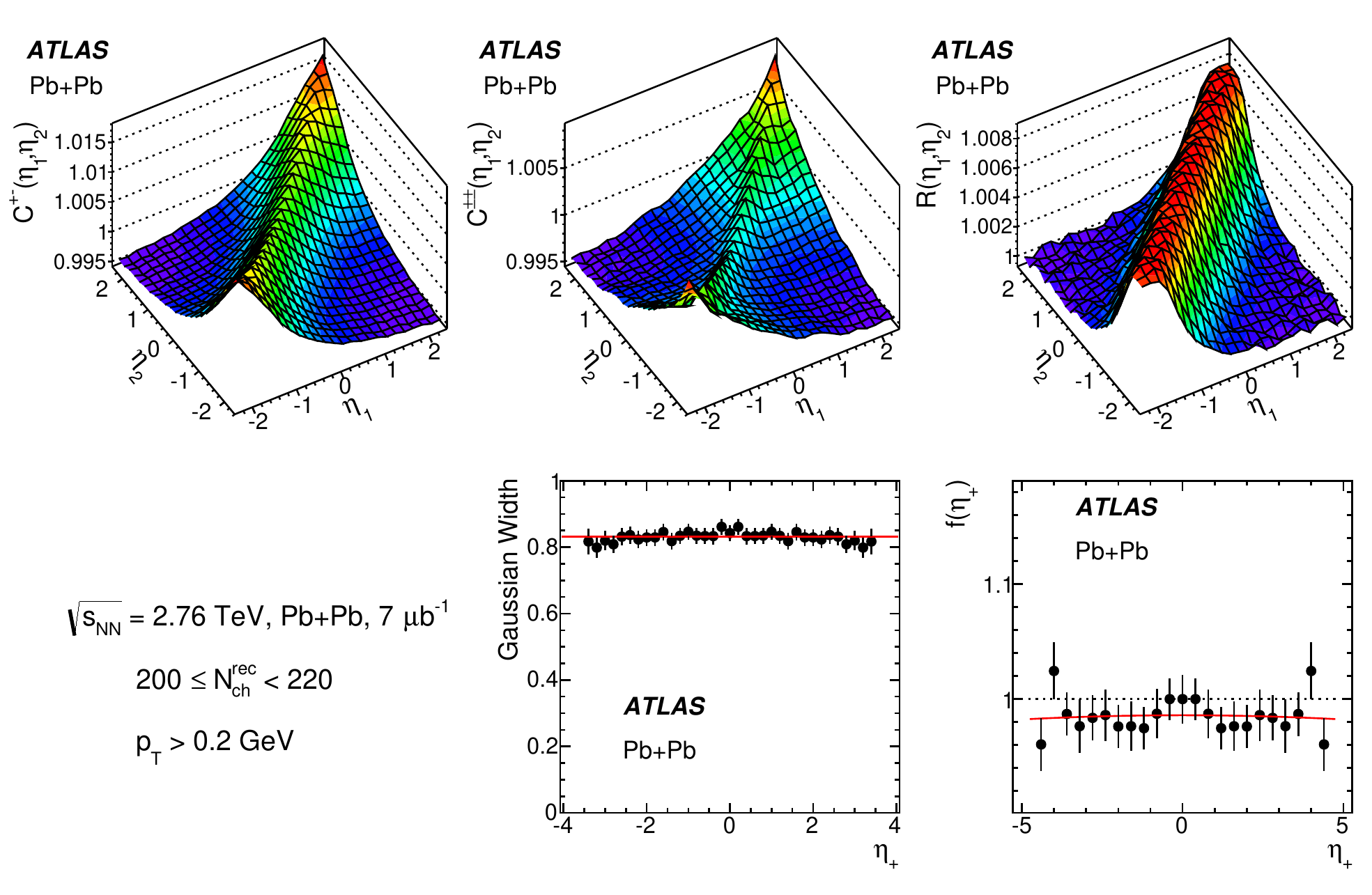}
\end{center}
\caption{\label{fig:1} The correlation functions for opposite-charge pairs $C^{+-}(\eta_1,\eta_2)$ (top-left panel), same-charge pairs $C^{\pm\pm}(\eta_1,\eta_2)$ (top-middle panel), and the ratio $R(\eta_1,\eta_2)= C^{+-}(\eta_1,\eta_2)/C^{\pm\pm}(\eta_1,\eta_2)$ (top-right panel) for Pb+Pb collisions with $200\leq\nchrec<220$. The width and magnitude of the short-range peak of the ratio are shown, as a function of $\eta_+$, in the lower-middle panel and lower-right panels, respectively. The error bars represent the statistical uncertainties, and the solid lines indicate a quadratic fit. The dotted line in the bottom-right panel serves to indicate better the deviation of $f(\eta_+)$ from 1.} 
\end{figure}

\begin{figure}[!t]
\begin{center}
\includegraphics[width=1\linewidth]{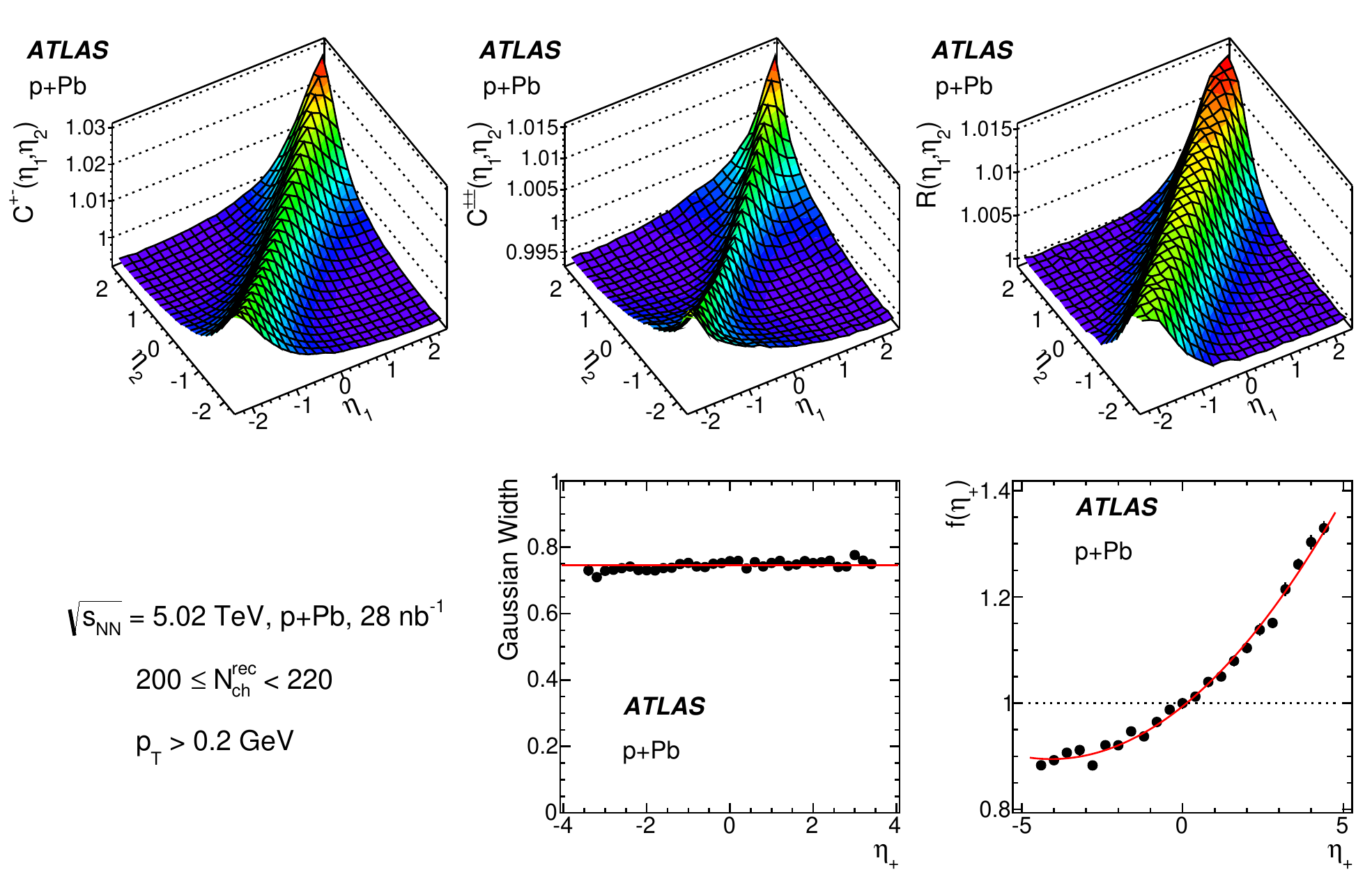}
\end{center}
\caption{\label{fig:2} The correlation functions for opposite-charge pairs $C^{+-}(\eta_1,\eta_2)$ (top-left panel), same-charge pairs $C^{\pm\pm}(\eta_1,\eta_2)$ (top-middle panel), and the ratio $R(\eta_1,\eta_2)= C^{+-}(\eta_1,\eta_2)/C^{\pm\pm}(\eta_1,\eta_2)$ (top-right panel) for $\pPb$ collisions with $200\leq\nchrec<220$. The width and magnitude of the short-range peak of the ratio are shown, as a function of $\eta_+$, in the lower-middle panel and lower-right panel, respectively. The error bars represent the statistical uncertainties, and the solid lines indicate a quadratic fit. The dotted line in the bottom-right panel serves to indicate better the deviation of $f(\eta_+)$ from 1.}
\end{figure}

\subsection{Probing the SRC via the same-charge and opposite-charge correlations}
\label{sec:34}
Figure~\ref{fig:1} shows separately the correlation functions for same-charge pairs and opposite-charge pairs from Pb+Pb collisions with $200\leq\nchrec<220$. The ratio of the two, $R(\eta_1,\eta_2)$ via Eq.~\eqref{eq:c0}, is shown in the top-right panel. The correlation functions show a narrow ``ridge''-like shape along $\eta_1\approx\eta_2$ or $\eta_-\approx 0$, and a falloff towards the corners at $\eta_1=-\eta_2\approx \pm2.4$. The magnitude of the ridge for the opposite-charge pairs is stronger than that for the same-charge pairs, which is characteristic of the influence from SRC from jet fragmentation or resonance decays. In regions away from the SRC, i.e. large values of $|\eta_{-}|$, the ratio approaches unity, suggesting that the magnitude of the LRC is independent of the charge combinations. To quantify the shape of the SRC in the ratio along $\eta_+$, $R$ is expressed in terms of $\eta_+$ and $\eta_-$, $R(\eta_+,\eta_-)$, and the following quantity is calculated:
\begin{eqnarray}
\label{eq:b1}
f(\eta_+) = \frac{\int_{-0.4}^{0.4} R(\eta_+,\eta_-)/0.8\;\,d\eta_--1}{\int_{-0.4}^{0.4} R(0,\eta_-)/0.8\;\,d\eta_--1}\;.
\end{eqnarray}
As shown in Fig.~\ref{fig:1}, the quantity $f(\eta_+)$ is nearly constant in Pb+Pb collisions, implying that the SRC is consistent with being independent of $\eta_+$. To quantify the shape of the SRC along the $\eta_-$ direction, $R(\eta_+,\eta_-)$ is fit to a Gaussian function in slices of $\eta_+$. The width, as shown in the bottom-middle panel of Fig.~\ref{fig:1}, is constant, which may suggest that the shape of the SRC in $\eta_-$ is the same for different $\eta_+$ slices.

Figure~\ref{fig:2} shows the correlation function in $\pPb$ collisions with multiplicity similar to the Pb+Pb data in Fig.~\ref{fig:1}. The correlation function shows a significant asymmetry between the proton-going side (positive $\eta_+$) and lead-going side (negative $\eta_+$). However, much of this asymmetry appears to be confined to a small $|\eta_-|$ region where the SRC dominates. The magnitude of the SRC, estimated by $f(\eta_+)$ shown in the bottom-right panel, increases by about 50\% from the lead-going side (negative $\eta_+$) to the proton-going side (positive $\eta_+$), but the width of the SRC in $\eta_-$ is independent of $\eta_+$ as shown in the bottom-middle panel. In contrast, the LRC has no dependence on the charge combinations, since the value of $R$ approaches unity at large $|\eta_-|$.

\begin{figure}[!]
\begin{center}
\includegraphics[width=0.6\linewidth]{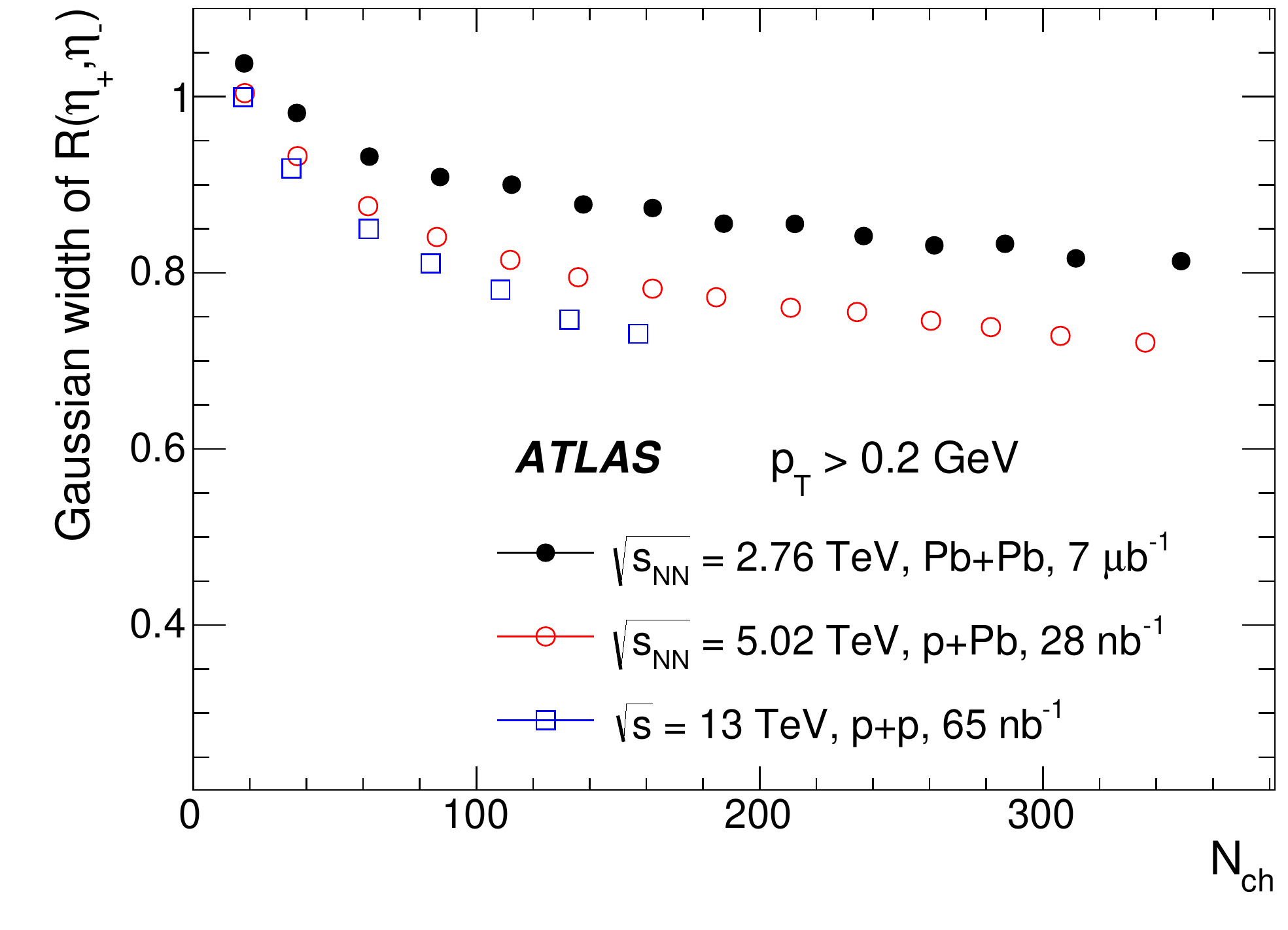}
\end{center}
\caption{\label{figaux:1} The width of the short-range component in $R(\eta_+,\eta_-)$ along the $\eta_-$ direction as a function of $\nch$ in the three collision systems.}
\end{figure}

\begin{figure}[!]
\begin{center}
\includegraphics[width=0.7\linewidth]{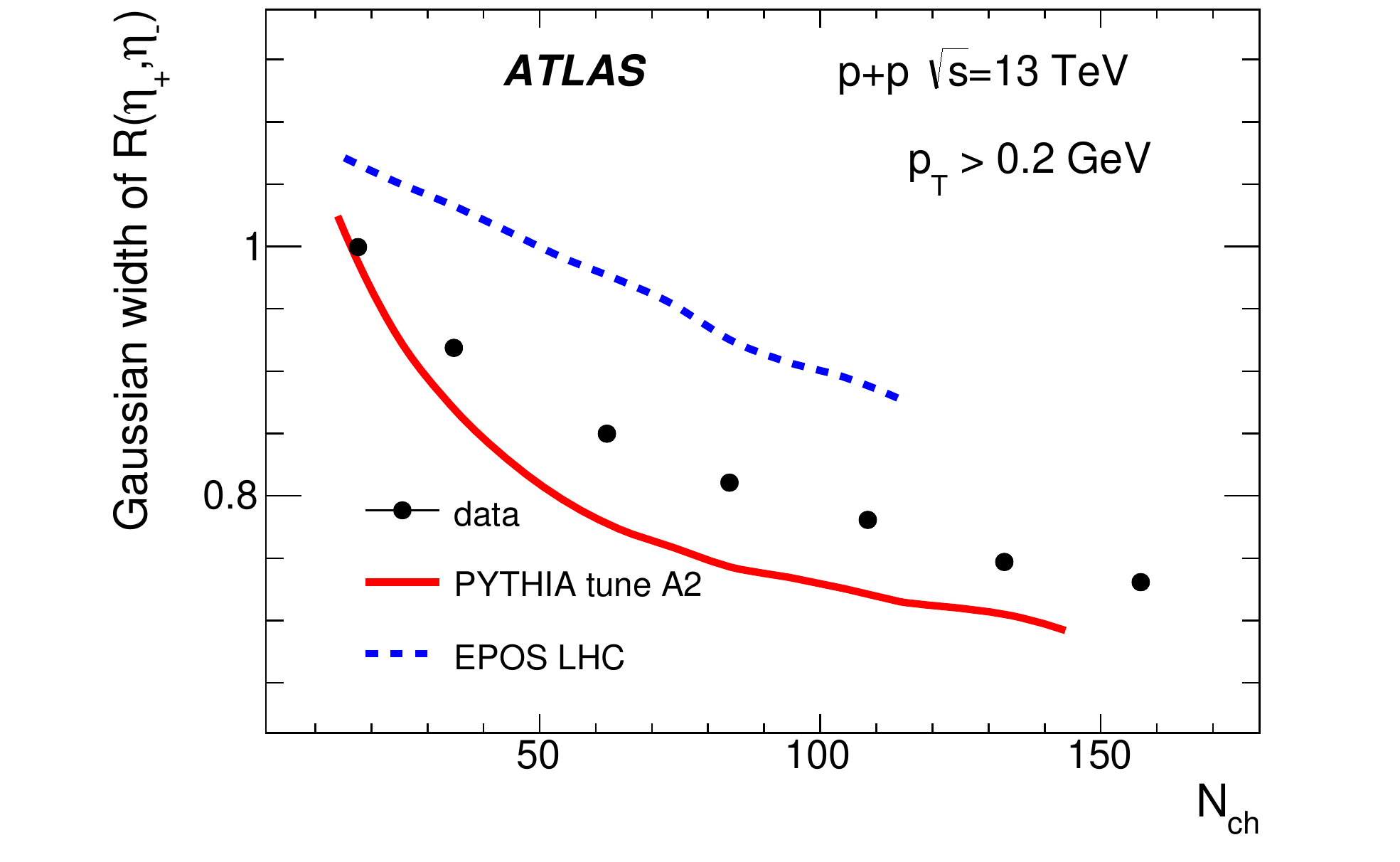}
\end{center}
\caption{\label{figaux:2} The width of the short-range component in $R(\eta_+,\eta_-)$ along the $\eta_-$ direction in $\pp$ collisions at $\sqrt{s}$ = 13 TeV, compared between data and two models. The y-axis is zero suppressed to demonstrate better the difference between data and models.}
\end{figure}

Figure~\ref{figaux:1} shows the width in $\eta_-$ of the short-range component as a function of $\nch$ in the three collision systems. The width is obtained as the Gaussian width of $R(\eta_+,\eta_-)$ in the $\eta_-$direction, and then averaged over $\eta_+$ as the width is observed to be independent of $\eta_+$, as shown in Figs.~\ref{fig:1} and \ref{fig:2}. This width reflects the extent of the short-range correlation in $\eta$, and it is observed to decrease with increasing $\nch$ in all collision systems. At the same $\nch$ value, the width is smallest in $\pp$ collisions and largest in Pb+Pb collisions. In Fig.~\ref{figaux:2}, the width of the short-range component from $\pp$ data is compared with PYTHIA 8 based on the A2 tune~\cite{PUB-2011-014} and EPOS based on the LHC tune~\cite{Pierog:2013ria}. The width is underestimated by PYTHIA 8 A2 and overestimated by EPOS LHC.

\subsection{Separation of the SRC and the LRC}
\label{sec:sub}

\begin{figure}[!b]
\begin{center}
\includegraphics[width=1\linewidth]{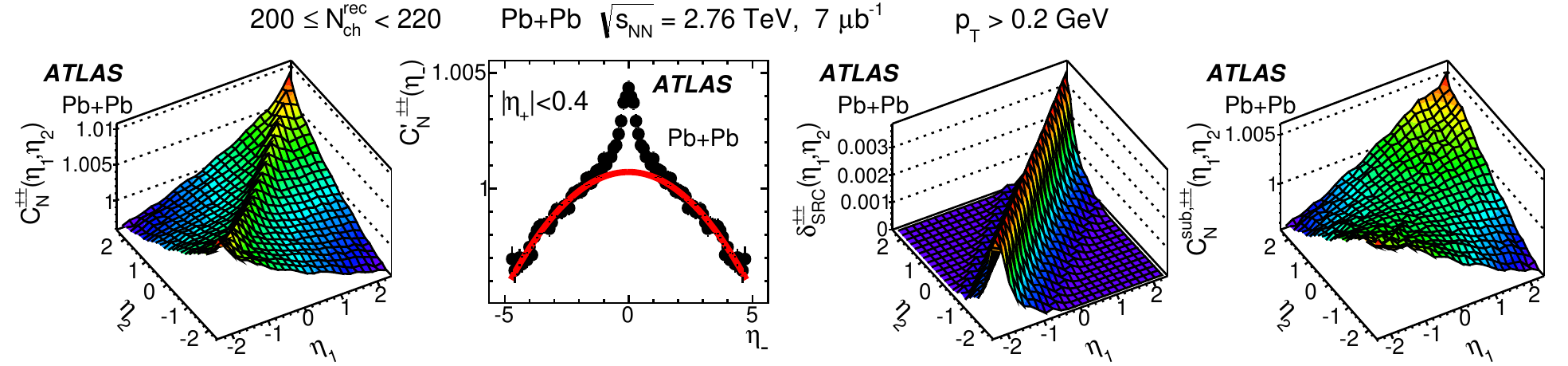}\\
\includegraphics[width=1\linewidth]{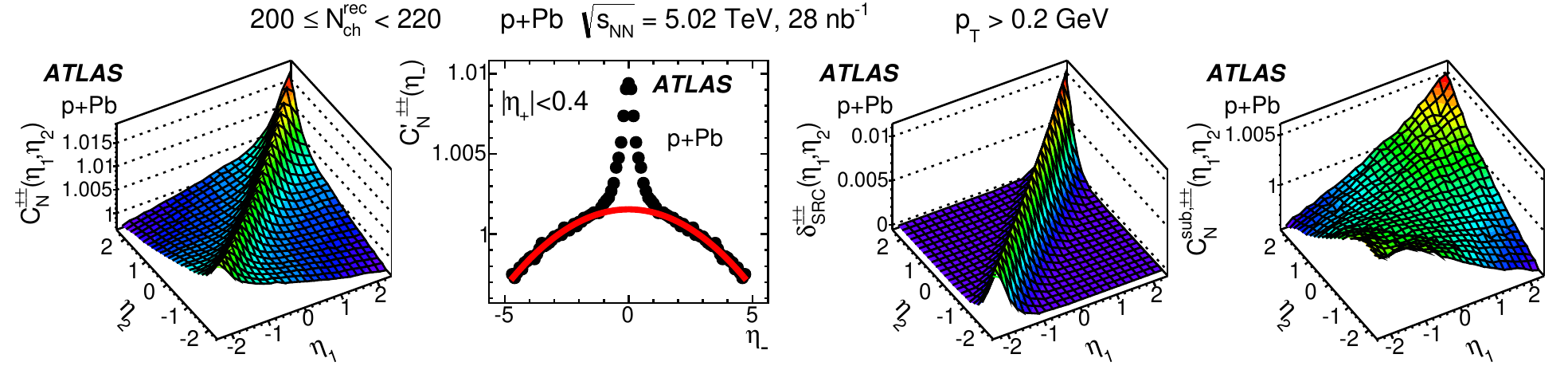}
\end{center}
\caption{\label{fig:4} The separation of correlation functions for same-charge pairs (first column) into the SRC (third column) and LRC (last column) for Pb+Pb (top row) and $\pPb$ (bottom row) collisions with $200\leq\nchrec<220$. The second column shows the result of the quadratic fit over the $|\eta_{-}|>1.5$ range of the 1-D correlation function projected over the $|\eta_{+}|<0.4$ slice, which is used to estimate the SRC component. The error bars represent the statistical uncertainties.}
\end{figure}

As discussed in Sec.~\ref{sec:33}, the ratio of the correlation function between opposite-charge and same-charge pairs $R(\eta_+,\eta_-)$ is the key to the separation of the SRC and LRC. Following Eqs.~\eqref{eq:c0} and \eqref{eq:c1}, this ratio can be approximated by:
\begin{eqnarray}
\label{eq:c2}
R(\eta_+,\eta_-) \approx 1 + f(\eta_+)\left[g^{+-}(\eta_-)-g^{\pm\pm}(\eta_-)\right] \;,\;\; \delta_{\mathrm{SRC}}^{+-}= f(\eta_+)g^{+-}(\eta_-),\;\; \delta_{\mathrm{SRC}}^{\pm\pm}= f(\eta_+)g^{\pm\pm}(\eta_-)
\end{eqnarray}
where $f(\eta_+)$ describes the shape along $\eta_+$ and can be calculated via Eq.~\eqref{eq:b1}. The functions $g^{+-}$ and $g^{\pm\pm}$ describe the SRC along the $\eta_-$ direction for the two charge combinations, which differ in both magnitude and shape.

In order to estimate the $g^{\pm\pm}(\eta_-)$ function for same-charged pairs, the $C_{\rm N}(\eta_+,\eta_-)$ distributions for same-charge pairs are projected into one-dimensional (1-D) $\eta_-$ distributions over a narrow slice $|\eta_+|<0.4$. The distributions are denoted by $C_{\rm N}(\eta_-)$. They are shown, after a small iterative correction discussed below, in the second column of Fig.~\ref{fig:4} for the same-charge pairs in Pb+Pb and $\pPb$ collisions. The SRC appears as a narrow peak on top of a distribution that has an approximately quadratic shape. Therefore a quadratic fit is applied to the data in the region of $|\eta_-|>1.5$, and the difference between the data and fit in the $|\eta_-|<2$ region is taken as the estimated SRC component or the $g^{\pm\pm}(\eta_-)$ function, which is assumed to be zero for $|\eta_-|>2$. This range ($|\eta_-|>1.5$) is about twice the width of the short-range peak in the $R(\eta_+,\eta_-)$ distribution along the $\eta_-$ direction (examples are given in the bottom-middle panel of Figures~\ref{fig:1} and \ref{fig:2}). This width is observed to decrease from 1.0 to 0.7 as a function of $\nchrec$ in the $\pPb$ collisions, and is slightly broader in Pb+Pb collisions and slightly narrower in $\pp$ collisions at the same $\nchrec$. The range of the fit is varied from $|\eta_-|>1.0$ to $|\eta_-|>2.0$ to check the sensitivity of the SRC estimation, and the variation is included in the final systematic uncertainties. Furthermore, this study is also repeated for $C_{\mathrm{N}}(\eta_-)$ obtained in several other $\eta_+$ slices within $|\eta_+|<1.2$, and consistent results are obtained. Once the distribution $g^{\pm\pm}(\eta_-)$ for same-charge pairs is obtained from the fit, it is multiplied by the $f(\eta_+)$ function calculated from $R(\eta_1,\eta_2)$ using Eq.~\eqref{eq:b1}, to obtain the $\delta_{\mathrm{SRC}}(\eta_1,\eta_2)$ from Eq.~\eqref{eq:c2} in the full phase space. Subtracting this distribution from the $C_{\rm N}(\eta_1,\eta_2)$ distribution, one obtains the initial estimate of the correlation function containing mostly the LRC component.

The LRC obtained via this procedure is still affected by a small bias from the SRC via the normalization procedure of Eq.~\eqref{eq:3}. This bias appears because the $\delta_{\mathrm{SRC}}(\eta_1,\eta_2)$ contribution is removed from the numerator but is still included in the denominator via $C_{p}(\eta)$. This contribution is not uniform in $\eta$: if the first particle is near mid-rapidity $\eta_1\approx0$ then all pairs in $\delta_{\mathrm{SRC}}(\eta_1,\eta_2)$ contribute to $C_p(\eta_1)$, whereas if the first particle is near the edge of the acceptance $\eta_1\approx\pm Y$ then only half of the pairs in $\delta_{\mathrm{SRC}}(\eta_1,\eta_2)$ contribute to $C_p(\eta_1)$. The acceptance bias in $C_{p}$ is removed via a simple iterative procedure: first, the $\delta_{\mathrm{SRC}}$ contribution determined from the above procedure is used to eliminate the SRC contribution to the single-particle mode:
\begin{eqnarray}
\label{eq:c2b}
C_{p}^{\rm{sub}}(\eta_1) = \frac{\int_{-Y}^{Y} \left[C(\eta_1,\eta_2)-\delta_{\mathrm{SRC}}(\eta_1,\eta_2)\right] \,d\eta_2}{2Y}\;,
\end{eqnarray}
with a similar expression for $C_{p}^{\rm{sub}}(\eta_2)$. The $C_{p}^{\rm{sub}}(\eta_1)$, $C_{p}^{\rm{sub}}(\eta_2)$ are then used to redefine the $C_{\mathrm{N}}$ function:
\begin{eqnarray}
\label{eq:c2c}
C_{\rm N}'(\eta_1,\eta_2) = \frac{C(\eta_1,\eta_2)}{C_{p}^{\rm{sub}}(\eta_1)C_{p}^{\rm{sub}}(\eta_2)}.
\end{eqnarray}
This distribution, which is very close to the distribution before correction, is shown in the second column of Fig.~\ref{fig:4} for projection over a narrow slice $|\eta_+|<0.4$. The estimation of $\delta_{\mathrm{SRC}}(\eta_1,\eta_2)$ is repeated using the previously described procedure for the $C_{\rm N}'(\eta_1,\eta_2)$, and the extracted distribution is shown in the third column of Fig.~\ref{fig:4}. Subtracting this distribution from $C_{\rm N}'(\eta_1,\eta_2)$, one obtains the correlation function containing only the LRC component. The resulting correlation function, denoted $C_{\rm N}^{\rm {sub}}(\eta_1,\eta_2)$, is shown in the last column of Fig.~\ref{fig:4}. 

The results presented in this paper are obtained using the iterative procedure discussed above. In most cases, the results obtained from the iterative procedure are consistent with the one obtained without iteration. In $\pPb$ and Pb+Pb collisions, where the SRC component is small, the difference between the two methods is found to be less than 2\%. In $\pp$ collisions with $\nchrec>100$, the difference between the two methods reaches 4\% where the SRC is large and therefore the bias correction is more important. 

In principle, the same analysis procedure can be applied to opposite-charge and all-charge pairs. However, due to the much larger SRC, the extracted LRC for opposite-charge pairs has larger uncertainties. Instead, the SRC for opposite-charge pairs is obtained directly by rearranging the terms in Eq.~\eqref{eq:c1} as:
\begin{eqnarray}
\label{eq:c1b}
\delta_{\mathrm{SRC}}^{+-}(\eta_1,\eta_2)  = R(\eta_1,\eta_2) -1 +\delta_{\mathrm{SRC}}^{\pm\pm}(\eta_1,\eta_2)\;.
\end{eqnarray}
The SRC for all-charge pairs is calculated as the average of $\delta_{\mathrm{SRC}}^{\pm\pm}$ and $\delta_{\mathrm{SRC}}^{+-}$ weighted by the number of same-charge and opposite-charge pairs. The LRC is then obtained by subtracting the SRC from the modified $C_{\rm N}(\eta_1,\eta_2)$ using the same procedure as that for the same-charge pairs.

For $\pp$ collisions, the pseudorapidity correlations are also compared with the PYTHIA 8 A2 and EPOS LHC event generators mentioned above. The analysis procedure used on the data is repeated for the two models in order to extract the SRC and LRC components. The correlation is carried out on the generated, as opposed to the reconstructed, charged particles.

\subsection{Quantifying the magnitude of the forward-backward multiplicity fluctuations}
\label{sec:a1}
In the azimuthal correlation analysis, the azimuthal structure of the correlation function is characterized by harmonic coefficients $v_n$ obtained via a Fourier decomposition~\cite{Aamodt:2011by,Aad:2012bu}. A similar approach can be applied for pseudorapidity correlations~\cite{Bzdak:2012tp,jjia}. Following Eq.~\eqref{eq:5}, the correlation functions are expanded into Legendre polynomial functions, and the two-particle Legendre coefficients $\lr{a_na_m}$ are calculated directly from the correlation function according to Eq.~\eqref{eq:6}. The two-particle correlation method measures, in effect, the RMS values of the EbyE $a_n$, and the final results for the coefficients are presented in terms of $\sqrt{|\lr{a_na_m}|}$. As a consequence of the condition for a symmetric collision system, the odd and even coefficients should be uncorrelated in $\pp$ and Pb+Pb collisions:
\begin{eqnarray}
a_{n,n+1}=\lr{a_na_{n+1}}=0\;.
\end{eqnarray}
However, even in $\pPb$ collisions, the correlation function after SRC removal, $C_{\rm N}^{\rm{sub}}(\eta_1,\eta_2)$, is observed to be nearly symmetric between $\eta$ and $-\eta$ (right column of Fig.~\ref{fig:4}), and hence the $\lr{a_na_{n+1}}$ values are very small and considered to be negligible in this paper.

\begin{figure}[!t]
\begin{center}
\includegraphics[width=0.7\linewidth]{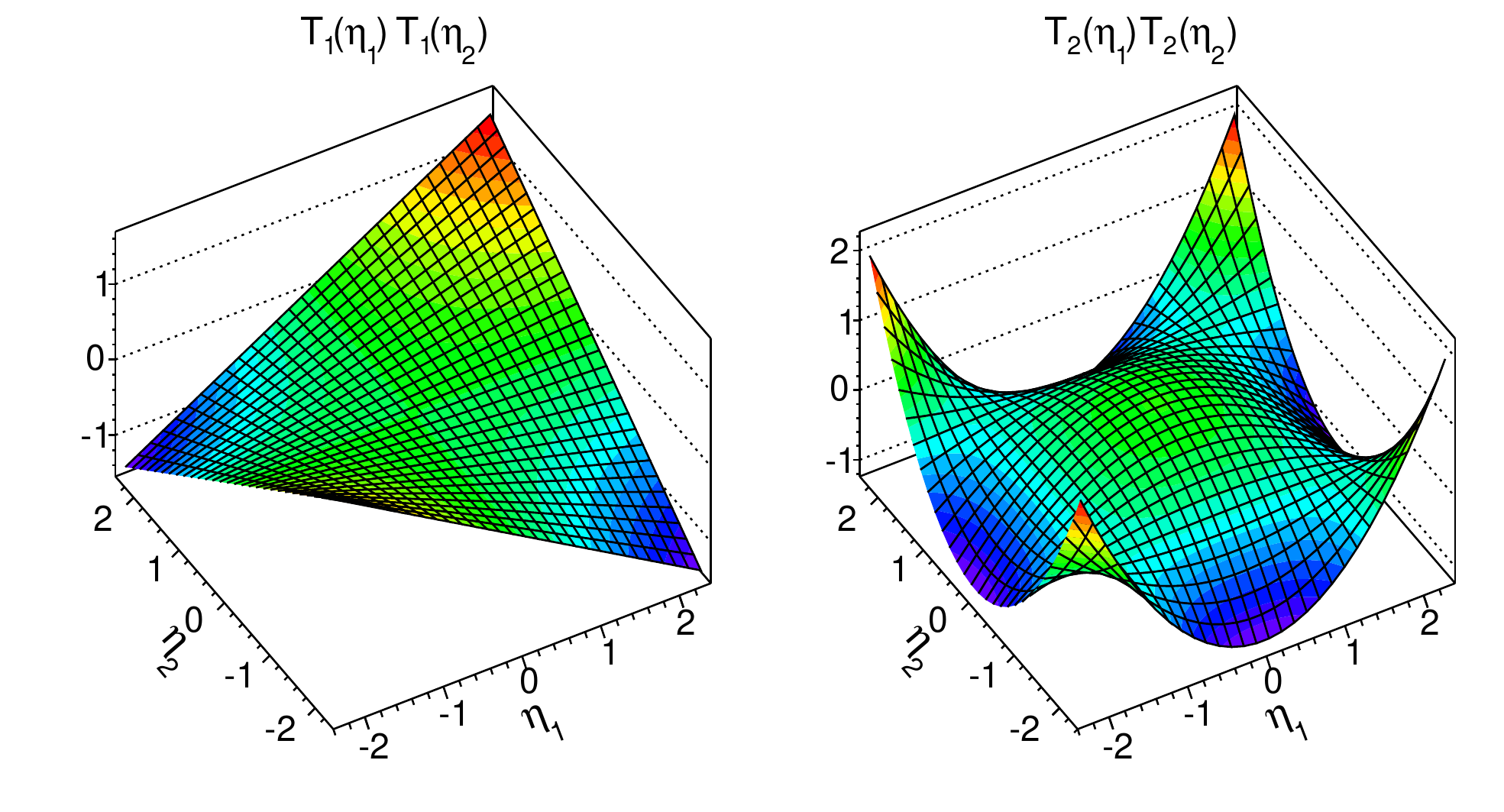}
\end{center}
\caption{\label{fig:d1} The first two Legendre basis functions associated with $a_{1,1}$ and $a_{2,2}$ in the two-particle correlation function.}
\end{figure}

The shape of the first two Legendre bases in 2-D are shown in Fig.~\ref{fig:d1}. The first basis function has the shape of $\eta_1\times\eta_2$ and is directly sensitive to the FB asymmetry of the EbyE fluctuation. The second basis function has a quadratic shape in the $\eta_1$ and $\eta_2$ directions and is sensitive to the EbyE fluctuation in the width of the $N(\eta)$ distribution. It is shown in Sec.~\ref{sec:result} that the data require only the first term, in which case the shape of the correlation function can be approximated by:
\begin{eqnarray}\label{eq:d1}
C_{\rm N}^{\rm{sub}}(\eta_1,\eta_2)\approx 1+ \lr{a_1^2}\eta_1\eta_2 =1+\frac{\lr{a_1^2}}{4}(\eta_+^2-\eta_-^2)\;.
\end{eqnarray}
Therefore a quadratic shape is expected along the two diagonal directions $\eta_+$ and $\eta_-$ of the correlation function, and the $\lr{a_1^2}^{\nicefrac{1}{2}}$ coefficient can be calculated by a simple quadratic fit of $C_{\rm N}^{\rm{sub}}$ in narrow slices of $\eta_-$ or $\eta_+$.

Alternatively, $\lr{a_1^2}^{\nicefrac{1}{2}}$ can also be estimated from a correlator constructed from a simple ratio:
\begin{eqnarray}\label{eq:d2}
r_{\rm N}^{\rm{sub}}(\eta,\eta_{\rm{ref}}) &=& \left\{\begin{array}{ll} 
C_{\rm N}^{\rm{sub}}(-\eta,\eta_{\rm{ref}})/{C_{\rm N}^{\rm{sub}}(\eta,\eta_{\rm{ref}})} & \;,\;\eta_{\rm{ref}}>0
\\
{C_{\rm N}^{\rm{sub}}(\eta,-\eta_{\rm{ref}})}/{C_{\rm N}^{\rm{sub}}(-\eta,-\eta_{\rm{ref}})}& \;,\;\eta_{\rm{ref}}<0
 \end{array}\right. \\
&\approx& 1-2\lr{a_1^2}\eta\eta_{\rm{ref}} \;,
\end{eqnarray}
where $\eta_{\rm{ref}}$ is a narrow interval of 0.2. This correlator has the advantage that most of the single-particle modes are even functions in $\eta$, so they cancel in the ratios. Therefore, this correlator provides a robust consistency check of any potential bias induced by the renormalization procedure of Eq.~\eqref{eq:3}. A similar quantity can also be calculated for $C_{\rm N}(\eta_1,\eta_2)$, denoted by $r_{\rm N}(\eta,\eta_{\rm{ref}})$.

In summary, this paper uses the following four different methods to estimate $\lr{a_1^2}^{\nicefrac{1}{2}}$:
\begin{enumerate}
\item
Legendre decomposition of the 2-D correlation function $C_{\rm N}^{\rm{sub}}(\eta_+,\eta_-)$, via Eq.~\eqref{eq:5}.
\item
Quadratic fit of $C_{\rm N}^{\rm{sub}}(\eta_-)$ in a narrow slice of $\eta_+$, which gives $\lr{a_1^2}^{\nicefrac{1}{2}}$ as a function of $\eta_+$.
\item
Quadratic fit of $C_{\rm N}^{\rm{sub}}(\eta_+)$ in a narrow slice of $\eta_-$, which gives $\lr{a_1^2}^{\nicefrac{1}{2}}$ as a function of $\eta_-$.
\item
Linear fit of $r_{\rm N}^{\rm{sub}}(\eta)$ in a narrow slice of $\eta_{\rm{ref}}$, which gives $\lr{a_1^2}^{\nicefrac{1}{2}}$ as a function of $\eta_{\rm{ref}}$.
\end{enumerate}
The three fitting methods (2,3,4) use the correlation function in limited and largely nonoverlapping regions of the $\eta_1$ and $\eta_2$ phase space, and therefore are independent of each other and largely independent of the Legendre decomposition method. Moreover, if the correlation function is dominated by the $\lr{a_1^2}$ term, the results from all four methods should be consistent.

\subsection{Systematic uncertainties}
\label{sec:syscheck}
The systematic uncertainties in this analysis arise from the event mixing, track reconstruction and selection efficiency, pair acceptance, and using simulated events to test the analysis process by comparing results from the generated charged particles with those from reconstructed tracks. These uncertainties apply to $C_{\rm N}(\eta_1, \eta_2)$ or $C_{\rm N}^{\rm{sub}}(\eta_1, \eta_2)$ and the associated Legendre coefficients. However, the systematic uncertainty for $C_{\rm N}^{\rm{sub}}(\eta_1, \eta_2)$ also depends on the procedure for separating the SRC from the LRC.

A natural way of quantifying these systematic uncertainties, used in this analysis, is to calculate $C_{\rm N}(\eta_1, \eta_2)$ or $C_{\rm N}^{\rm{sub}}(\eta_1, \eta_2)$ under a different condition, and then construct the ratio to the default analysis: $D(\eta_1, \eta_2)$. The average deviation of $D(\eta_1, \eta_2)$ from unity can be compared with the correlation signal to estimate the systematic uncertainties in the correlation function. The same $D(\eta_1, \eta_2)$ function can also be expanded into a Legendre series (Eq.~\eqref{eq:5}), and the resulting coefficients $a_{n,m}^{\mathrm{d}}$ can be used to estimate the systematic uncertainties for the $a_{n,m}$ coefficients. For the three fitting methods discussed in Sec.~\ref{sec:a1}, the fits are repeated for each check to estimate the uncertainties in the resulting $\lr{a_1^2}^{\nicefrac{1}{2}}$ values. These uncertainties are not always the same for $C_{\rm N}$ and $C_{\rm N}^{\rm{sub}}$ because $C_{\rm N}^{\rm{sub}}$ is not sensitive to the variation in the short-range region, $\eta_-\approx0$. In the following, the uncertainty from each source is discussed.

The main source of uncertainty for $C_{\rm N}^{\rm{sub}}(\eta_1, \eta_2)$ arises from the procedure to separate the SRC and the LRC. Since the estimated SRC component for the opposite-charge pairs is more than a factor of two larger than that for the same-charge pairs (e.g. Figs.~\ref{fig:1}--\ref{fig:2}), the difference between $C_{\rm N}^{\rm{sub,+-}}$ and $C_{\rm N}^{\rm{sub,\pm\pm}}$ is a conservative check of the robustness of the subtraction procedure. This difference is typically small for events with large $\nchrec$, and it is found to be within 0.2--2.2\% of the correlation signal and 1--6\% for $\lr{a_1^2}^{\nicefrac{1}{2}}$ in the three collision systems. The stability of LRC is also checked by varying the fit range and varying the $\eta_+$ slice used to obtain the $\delta_{\mathrm{SRC}}(\eta_-)$ distribution for same-charge pairs. This uncertainty amounts to 1--2\% in the correlation signal and 1--5\% for $\lr{a_1^2}^{\nicefrac{1}{2}}$ in Pb+Pb collisions, and is larger in $\pPb$ and $\pp$ collisions due to a stronger SRC for events with the same $\nchrec$. 

Uncertainties due to the event-mixing are evaluated by varying the criteria for matching events in $\nchrec$ and $z_{\mathrm{vtx}}$. The $a_{n,m}^{\mathrm{d}}$ values are calculated for each case. The uncertainty from variation of the matching range in $z_{\mathrm{vtx}}$ is less than 0.5\% of the correlation signal for both $C_{\rm N}$ and $C_{\rm N}^{\rm{sub}}$. The bin size in $\nchrec$ for event matching is varied such that the number of events in each bin varies by a factor of three. Most of the changes appear as modulations of the projections of the correlation function in $\eta_1$ or $\eta_2$ as defined in Eq.~\eqref{eq:4}, and the renormalized correlation functions $C_{\rm N}(\eta_1, \eta_2)$ and $C_{\rm N}^{\rm{sub}}(\eta_1, \eta_2)$ are very stable. The difference between different variations amounts to at most 2\% of the correlation signal or $\lr{a_1^2}^{\nicefrac{1}{2}}$. The analysis is also repeated separately for events with $|z_{\mathrm{vtx}}|<50$~mm and $50<|z_{\mathrm{vtx}}|<100$~mm. Good agreement is seen between the two. To evaluate the stability of the correlation function, the entire dataset is divided into several groups of runs, and the correlation functions and $a_n$ coefficients are calculated for each group. The results are found to be consistent within 2\% for $\lr{a_1^2}^{\nicefrac{1}{2}}$.

The 13~$\TeV$ $\pp$ results are obtained from the June 2015 and August 2015 datasets with different $\mu$ values. The influence of the residual pileup is evaluated by comparing the results obtained separately from these two running periods, and no systematic difference is observed between the results.

The shape of the correlation function is not very sensitive to the uncertainty in the tracking efficiency correction, since this correction is applied in both the numerator and denominator. On the other hand, both the correlation signal and reconstruction efficiency are observed to increase with $\pT$, and hence the correlation signal and associated $\lr{a_na_m}$ coefficients are expected to be smaller when corrected for reconstruction efficiency. Indeed, a 1--2\% decrease in $\lr{a_n^2}^{\nicefrac{1}{2}}$ is observed after applying this correction. This change is conservatively included in the systematic uncertainty.

The correlation function $C_{\rm N}(\eta_1,\eta_2)$ has some small localized structures that are not compatible with statistical fluctuations. These structures are due to residual detector effects in the pair acceptance that are not removed by the event-mixing procedure, which can be important for extraction of the higher-order coefficients. Indeed, the Legendre coefficients for $n\geq8$ show significant nonstatistical fluctuations around zero. Therefore, the spread of $\lr{a_n^2}^{\nicefrac{1}{2}}$ for $n\geq10$ and $\sqrt{|\lr{a_na_{n+2}}|}$ for $n\geq8$ are quoted as uncertainties for the Legendre coefficients. These uncertainties are less than $0.5\times10^{-5}$ for $\lr{a_na_m}$ calculated from $C_{\rm N}^{\rm {sub}}(\eta_1, \eta_2)$ in all collision systems, and are larger for those calculated from $C_{\rm N}(\eta_1, \eta_2)$. The corresponding relative uncertainty for $\lr{a_1^2}$ is negligible.

The HIJING and PYTHIA 8 events used for evaluating the reconstruction efficiency have a significant correlation signal and sizable $a_{n,m}$ coefficients for $C_{\rm N}$. The correlation functions obtained using the reconstructed tracks are compared with those obtained using the generated charged particles. The ratio of the two is then used to vary the measured $C_{\rm N}(\eta_1,\eta_2)$, the procedure for removal of the SRC is repeated and the variations of $C_{\rm N}^{\rm {sub}}$ and $a_{n,m}$ are calculated. The differences in the correlation function reflect mainly the uncertainty in the efficiency correction, but also the influence of secondary decays and fake tracks. These differences are found to be mostly concentrated in a region around $\eta_-\approx0$, and hence affect mostly the estimation of the SRC component, and have very little impact on $C_{\rm N}^{\rm {sub}}$ and associated $a_{n,m}$. The differences in Legendre coefficients are found to be up to 5\% for $a_n$ calculated from $C_{\rm N}$, and are 0.2--3.5\% for $\lr{a_1^2}^{\nicefrac{1}{2}}$ calculated from $C_{\rm N}^{\rm {sub}}$.
\vspace*{-0.3cm}
\begin{table}[!t]
\centering
\caption{Summary of average systematic uncertainties for the correlation function $C_{\rm N}^{\rm {sub}}(\eta_1,\eta_2)$ with $\pT>0.2$~$\GeV$. The uncertainty is calculated as the variation relative to the correlation signal of $C_{\rm N}^{\rm {sub}}(\eta_1,\eta_2)$, averaged over the entire $\eta_1$ and $\eta_2$ space. The range in the table covers the variation of this uncertainty for different $\nchrec$ classes.}\vspace*{0.3cm}
\label{tab:corr}
{\begin{tabular}{|l|ccc|}\hline 
Collision system                  &   Pb+Pb       &   $\pPb$     &  $\pp$       \tabularnewline\hline
Charge dependence             [\%]&   0.2--1.6    &  0.2--1.9   & 0.7--2.2     \tabularnewline
SRC LRC separation               [\%]&   1.0--2.2    &  1.2--5.7   & 1.1--3.9     \tabularnewline
Event-mixing                  [\%]&   0.7--1.0    &   0.4--2.5  & 0.2--1.8     \tabularnewline
$\zvtx$ variation             [\%]&   0.4--0.7    &   0.3--1.8  & 0.2--2.0     \tabularnewline
Run-by-run stability          [\%]&   0.4--0.8    &   0.3--1.7  & 0.2--1.6     \tabularnewline
Track selection \& efficiency [\%]&   0.7--1.4    &   0.2--0.3  & 0.3--0.6     \tabularnewline
MC consistency                [\%]&   0.4--2.2    &   0.6--2.9  & 0.6--2.9     \tabularnewline
& & & \tabularnewline\hline
Total [\%]                        &   1.6--3.6    &  1.6--7.2   & 2.0--5.9 \tabularnewline\hline\hline
\end{tabular}}
\end{table}
\begin{table}[!t]
\centering
\vspace*{-0.3cm}
\caption{Summary of systematic uncertainties for $\lr{a_1^2}^{\nicefrac{1}{2}}$ with $\pT>0.2$~$\GeV$, calculated with four different methods: Legendre expansion of $C_{\rm N}^{\rm {sub}}(\eta_1,\eta_2)$, quadratic fit of the $\eta_-$ dependence of $C_{\rm N}^{\rm{sub}}(\eta_1,\eta_2)$ for $|\eta_+|<0.1$, quadratic fit of the $\eta_+$ dependence of $C_{\rm N}^{\rm{sub}}(\eta_1,\eta_2)$ for $0.9<|\eta_-|<1.1$, and linear fit of the $\eta$ dependence of $r_{\rm N}^{\rm{sub}}(\eta,\eta_{\rm{ref}})$ for $2.2<|\eta_{\rm{ref}}|<2.4$.}
\label{tab:a1}
\vspace*{0.3cm}
\small{\begin{tabular}{|l|ccc|ccc|}\hline 
 &\multicolumn{3}{|c|}{ $\vphantom{\int^A_|}{}$ Quadratic fit to $C_{\rm N}^{\rm{sub}}(\eta_-)|_{|\eta_+|<0.1}$}& \multicolumn{3}{|c|}{Quadratic fit to the $C_{\rm N}^{\rm{sub}}(\eta_+)|_{0.9<|\eta_-|<1.1}$} \tabularnewline\hline
Collision system             &   Pb+Pb     &   $\pPb$    &  $\pp$     &   Pb+Pb       & $\pPb$   &  $\pp$  \tabularnewline\hline
Charge dependence        [\%]& 0.1--2.7    & 0.4--2.5    & 1.1--3.4   &  0.2--5.5     & 0.5--7.0 &  1.2--7.3  \tabularnewline
SRC LRC separation          [\%]& 1.2--2.6    & 1.1--6.7    & 1.4--5.3   &  1.0--2.9     & 0.8--3.1 &  1.8--3.5  \tabularnewline
Event-mixing             [\%]& 0.5--2.5    & 0.2--2.8    & 0.2--4.2   &  0.4--1.8     & 0.4--3.2 &  0.3--3.4  \tabularnewline
$\zvtx$ variation        [\%]& 0.4--2.2    & 0.2--1.5    & 0.2--1.4   &  0.3--1.7     & 0.2--2.4 &  0.2--3.7  \tabularnewline
Run-by-run stability     [\%]& 0.3--2.1    & 0.2--1.8    & 0.2--3.0   &  0.2--2.4     & 0.2--2.1 &  0.2--1.5  \tabularnewline
Track selec.\& efficiency[\%]& 0.6--4.4    & 0.5--1.0    & 1.0--1.9   &  0.7--4.7     & 0.7--1.0 &  0.8--1.4   \tabularnewline
MC consistency           [\%]& 0.5--4.5    & 0.4--4.9    & 1.8--7.2   &  0.8--5.1     & 0.2--5.8 &  0.4--8.1  \tabularnewline
& & & & & &\tabularnewline
Total [\%]                   & 2.1--6.2    & 1.8--7.5   & 3.1--9.7    &  2.2--5.6     & 1.9--6.2 &  2.8--10.0  \tabularnewline\hline\hline
 &\multicolumn{3}{|c|}{ $\vphantom{\int^A_|}{}$ Linear fit to $r_{\rm N}^{\rm{sub}}(\eta)|_{2.2<|\eta_{\rm{ref}}|<2.4}$}   & \multicolumn{3}{|c|}{Global Legendre expansion of $C_{\rm N}^{\rm{sub}}$} \tabularnewline\hline
Collision system             &   Pb+Pb  &   $\pPb$ &  $\pp$      &   Pb+Pb     &   $\pPb$ &  $\pp$  \tabularnewline\hline
Charge dependence        [\%]& 0.3--3.4 & 0.4--3.5 &  0.9--4.3   & 0.3--4.5    & 0.4--5.2 &  1.5--6.3 \tabularnewline
SRC LRC separation          [\%]& 1.3--2.4 & 1.2--2.4 &  1.4--2.7   & 1.2--4.5    & 2.2--8.8 &  2.5--5.9 \tabularnewline
Event-mixing             [\%]& 0.4--2.2 & 0.4--1.2 &  0.3--2.6   & 0.2--1.7    & 0.2--1.6 &  0.2--0.4 \tabularnewline
$\zvtx$ variation        [\%]& 0.2--1.6 & 0.2--2.6 &  0.2--2.7   & 0.2--1.7    & 0.2--2.8 &  0.2--2.5 \tabularnewline
Run-by-run stability     [\%]& 0.2--1.9 & 0.1--2.2 &  0.2--3.0   & 0.2--0.6    & 0.1--1.8 &  0.2--2.2 \tabularnewline
Track selec.\& efficiency[\%]& 0.6--2.2 & 0.3--1.0 &  1.0--1.5   & 0.5--1.4    & 0.5--1.0 &  1.1--2.1 \tabularnewline
MC consistency           [\%]& 0.6--4.4 & 0.2--4.8 &  0.8--3.4   & 0.5--4.3    & 0.8--4.6 &  0.2--4.0 \tabularnewline
& & & & & &\tabularnewline
Total [\%]                   & 2.4--4.9 & 1.8--5.3 &  2.4--4.5   & 2.3--5.0    & 2.5--9.1 &  3.4--8.2 \tabularnewline\hline\hline
\end{tabular}}\normalsize
\end{table}

The systematic uncertainties from the different sources described above are added in quadrature to give the total systematic uncertainties for the correlation functions and $\lr{a_1^2}^{\nicefrac{1}{2}}$ values for both $C_{\rm N}$ and $C_{\rm N}^{\rm {sub}}$. The systematic uncertainties associated with $C_{\rm N}^{\rm {sub}}(\eta_1,\eta_2)$ and $\lr{a_1^2}^{\nicefrac{1}{2}}$ are given in Tables~\ref{tab:corr} and \ref{tab:a1}, respectively. Since there are four methods for extracting $\lr{a_1^2}^{\nicefrac{1}{2}}$, they are given separately in Table~\ref{tab:a1}. The systematic uncertainty quoted for each source in both tables covers the maximum uncertainty in the specified collision system. 

\section{Results}
\label{sec:result}
\begin{figure}[!b]
\begin{center}
\includegraphics[width=1\linewidth]{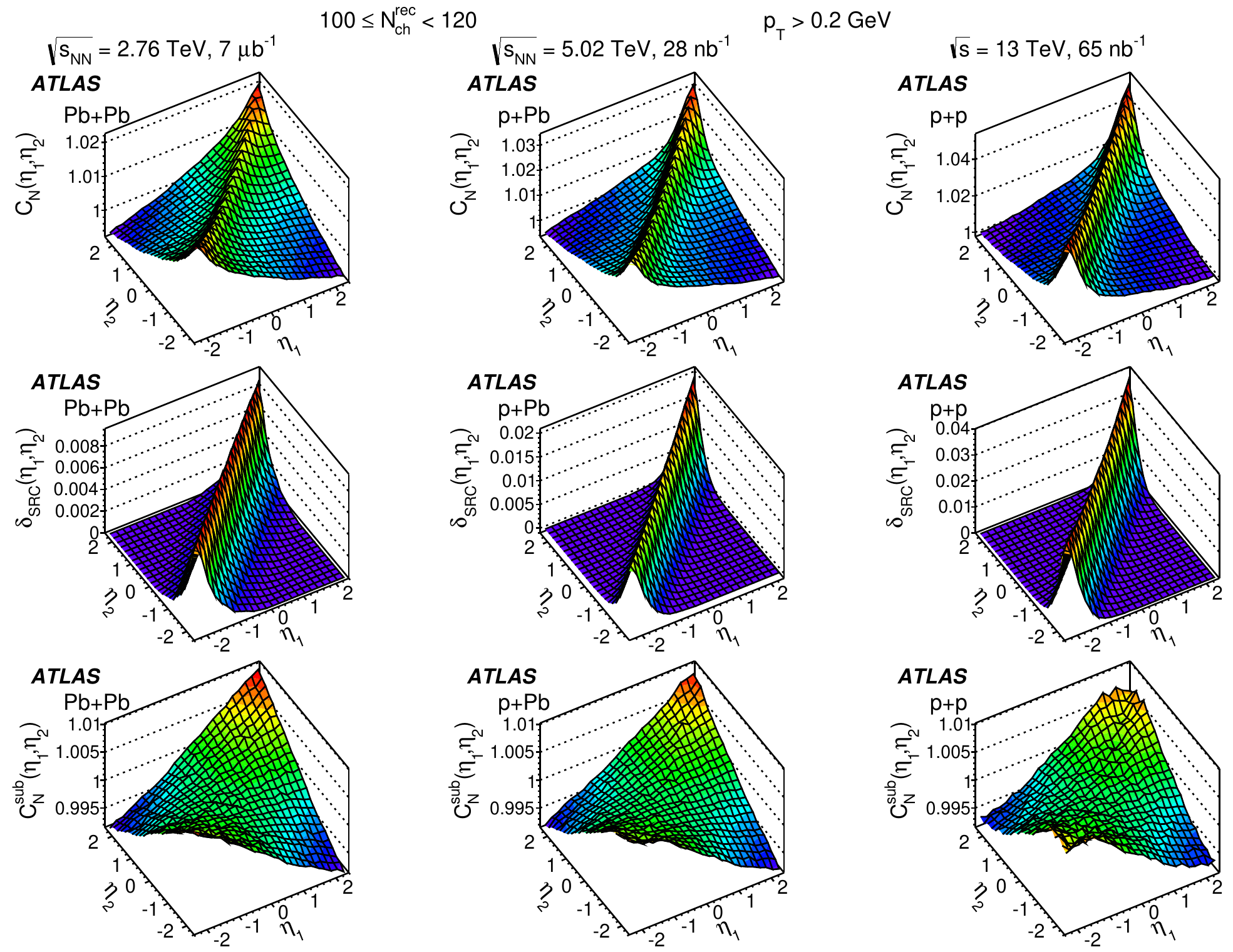}
\end{center}
\caption{\label{fig:5} The distributions of correlation functions $C_{\rm N}(\eta_1,\eta_2)$ (top row), the estimated short-range component $\delta_{\rm {SRC}}(\eta_1,\eta_2)$ (middle row), and long-range component $C_{\rm N}^{\rm {sub}}(\eta_1,\eta_2)$ (bottom row). They are shown for collisions with $100\leq\nchrec<120$ in Pb+Pb (left column), $\pPb$ (middle column), and $\pp$ collisions (right column).}
\end{figure}

The top row of Fig.~\ref{fig:5} shows the correlation functions $C_{\rm N}(\eta_1,\eta_2)$ in the three collision systems for events with similar multiplicity $100\leq\nchrec<120$. The corresponding estimated SRC component $\delta_{\rm {SRC}}(\eta_1,\eta_2)$ and long-range component $C_{\rm N}^{\rm {sub}}(\eta_1,\eta_2)$ are shown in the middle and bottom rows, respectively. The magnitude of the SRC in $\pPb$ is observed to be larger in the proton-going direction than in the lead-going direction, reflecting the fact that the particle multiplicity is smaller in the proton-going direction. However, this forward-backward asymmetry in $\pPb$ collisions is mainly associated with the SRC component, and the $C_{\rm N}^{\rm {sub}}(\eta_1,\eta_2)$ distribution shows very little asymmetry. The $C_{\rm N}(\eta_1,\eta_2)$ distributions show significant differences between the three systems, which is mainly due to their differences in $\delta_{\rm {SRC}}(\eta_1,\eta_2)$.  In fact the estimated long-range component $C_{\rm N}^{\rm {sub}}(\eta_1,\eta_2)$ shows similar shape and similar overall magnitude for the three systems.

\begin{figure}[!t]
\begin{center}
\includegraphics[width=0.33\linewidth]{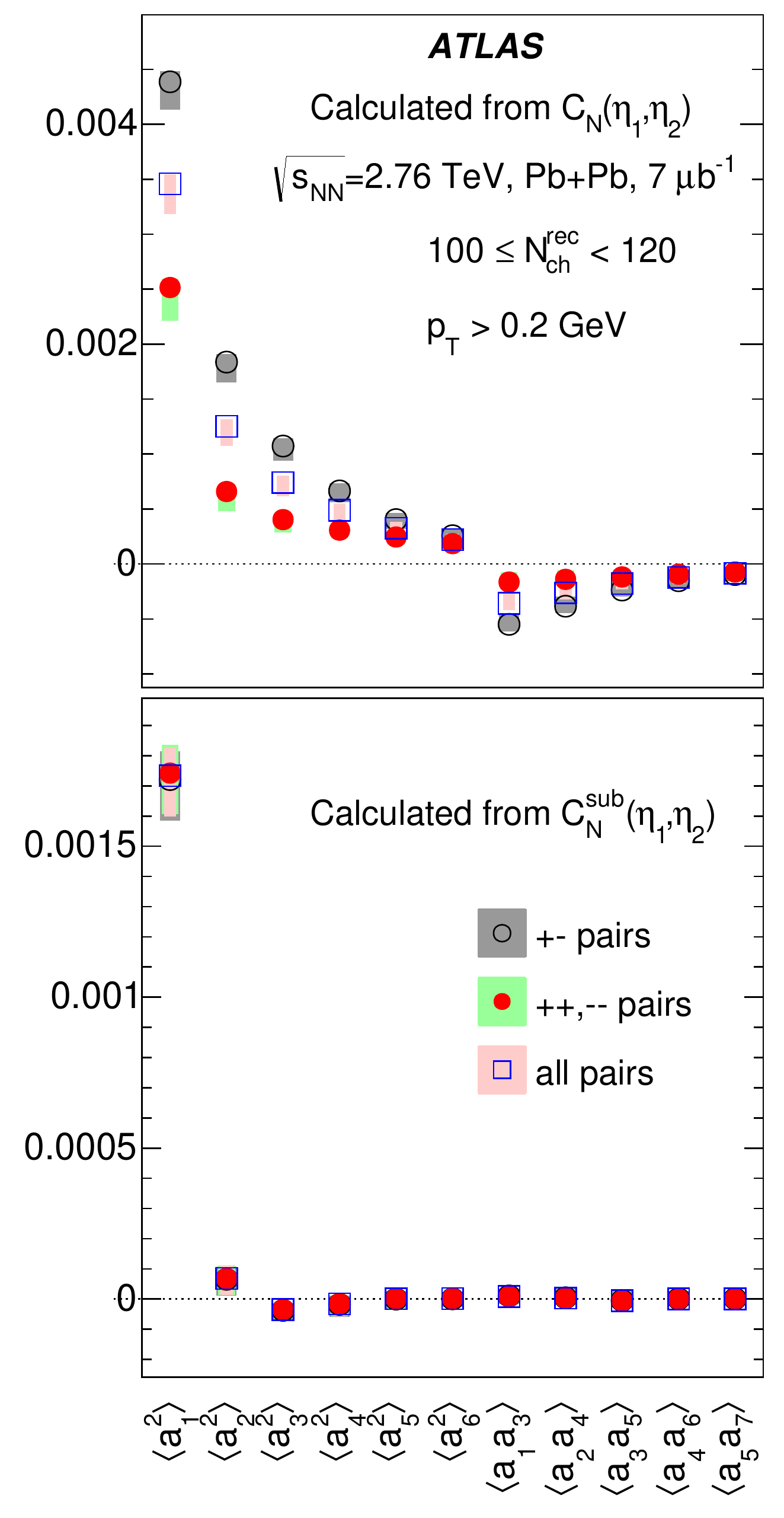}\includegraphics[width=0.33\linewidth]{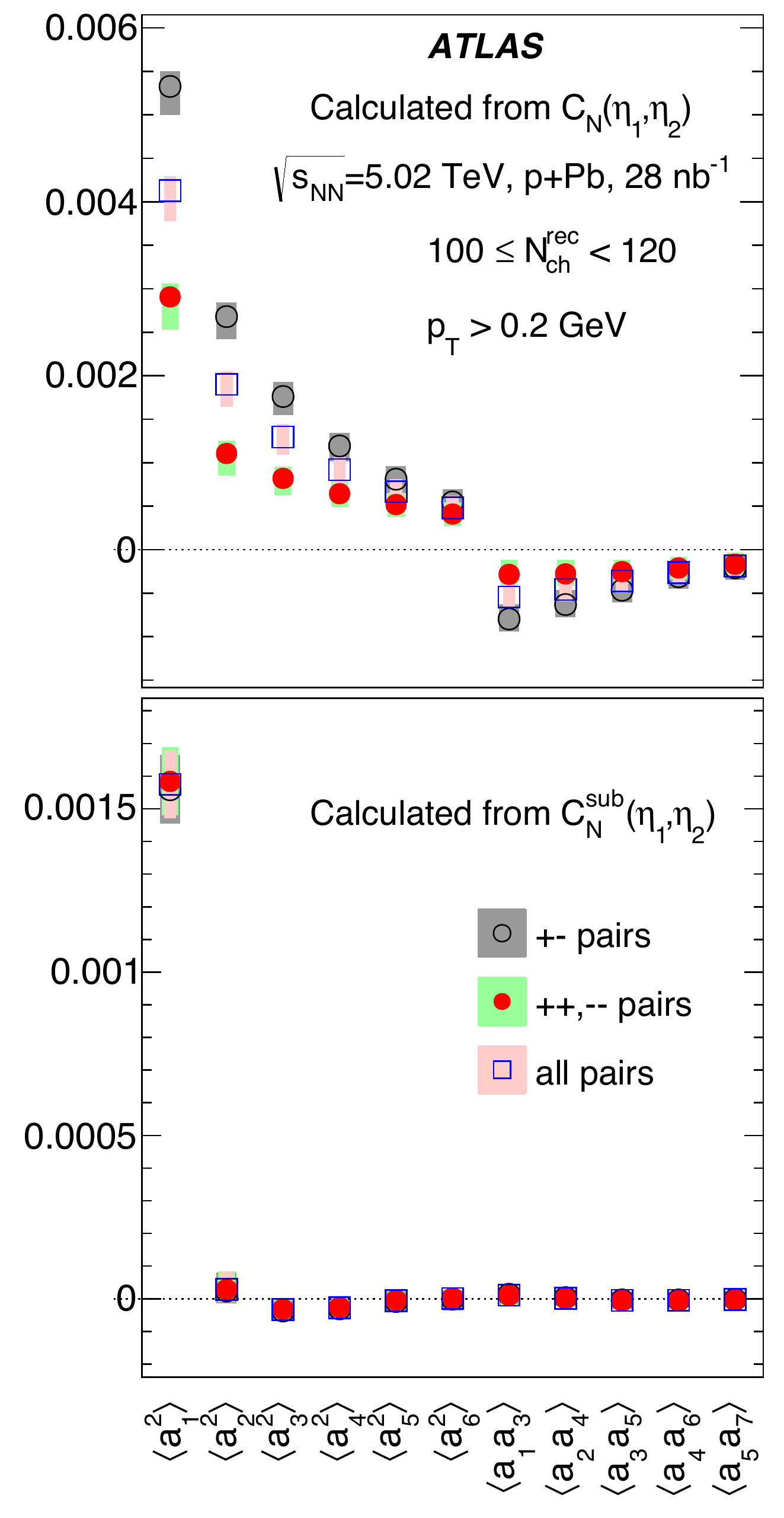}\includegraphics[width=0.33\linewidth]{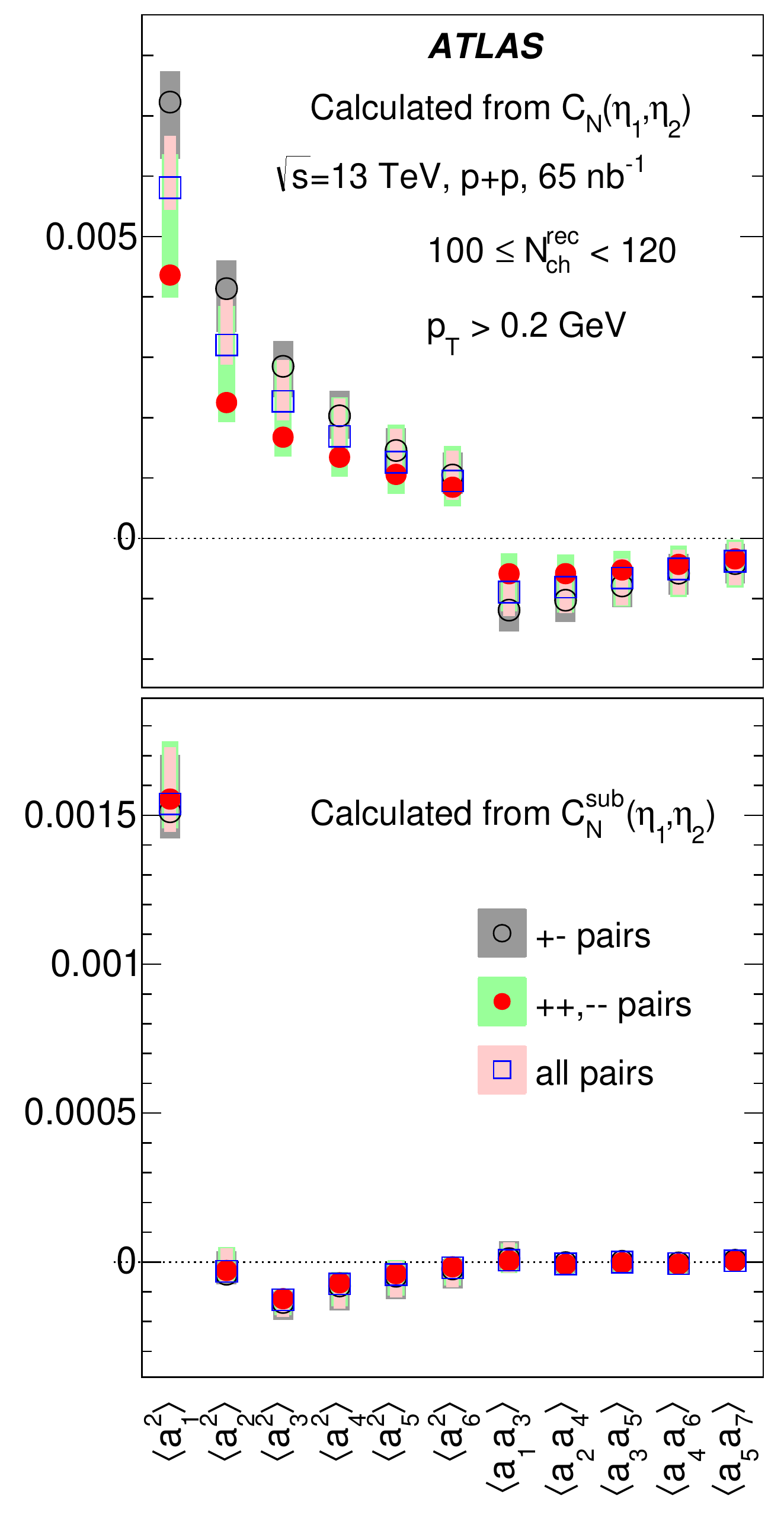}
\end{center}
\caption{\label{fig:6} The Legendre spectra $\lr{a_n^2}$ and $\lr{a_na_{n+2}}$ calculated via Eq.~\eqref{eq:6} from correlation functions $C_{\rm N}(\eta_1,\eta_2)$ (top row) and $C_{\rm N}^{\rm {sub}}(\eta_1,\eta_2)$ (bottom row) in Pb+Pb (left column), $\pPb$ (middle column), and $\pp$ (right column) collisions for events with $100\leq\nchrec<120$. The shaded bands represent the total uncertainties. The results are shown for all-charge (open squares), opposite-charge (open circles), and same-charge pairs (solid circles).}
\end{figure}

To characterize the shape of the correlation functions, the Legendre coefficients $\lr{a_na_m}$ for the distributions $C_{\rm N}$ and $C_{\rm N}^{\rm {sub}}$ shown in Fig.~\ref{fig:5} are calculated via Eq.~\eqref{eq:6} and plotted in Fig.~\ref{fig:6}. The $\lr{a_na_m}$ values are shown for the first six diagonal terms $\lr{a_n^2}$ and the first five mixed terms $\lr{a_na_{n+2}}$, and they are also compared with coefficients calculated for opposite-charge pairs and same-charge pairs for the same event class. The magnitudes of the $\lr{a_na_m}$ coefficients calculated for $C_{\rm N}$ differ significantly for the different charge combinations, and they also increase as the size of the collision system decreases, i.e. $|\lr{a_na_m}|_{p\mathrm{+}p}>|\lr{a_na_m}|_{p\mathrm{+Pb}}>|\lr{a_na_m}|_{\mathrm{Pb+Pb}}$. This is consistent with a large contribution from SRC to all $\lr{a_na_m}$ coefficients obtained from $C_{\rm N}$. After removal of the SRC, the $\lr{a_1^2}$ coefficient is quite consistent between different charge combinations and different collision systems. All higher-order coefficients are much smaller, and they are very close to zero within the systematic uncertainties. Therefore, the rest of the paper focuses on the $\lr{a_1^2}^{\nicefrac{1}{2}}$ results.

\begin{figure}[!t]
\begin{center}
\includegraphics[width=1\linewidth]{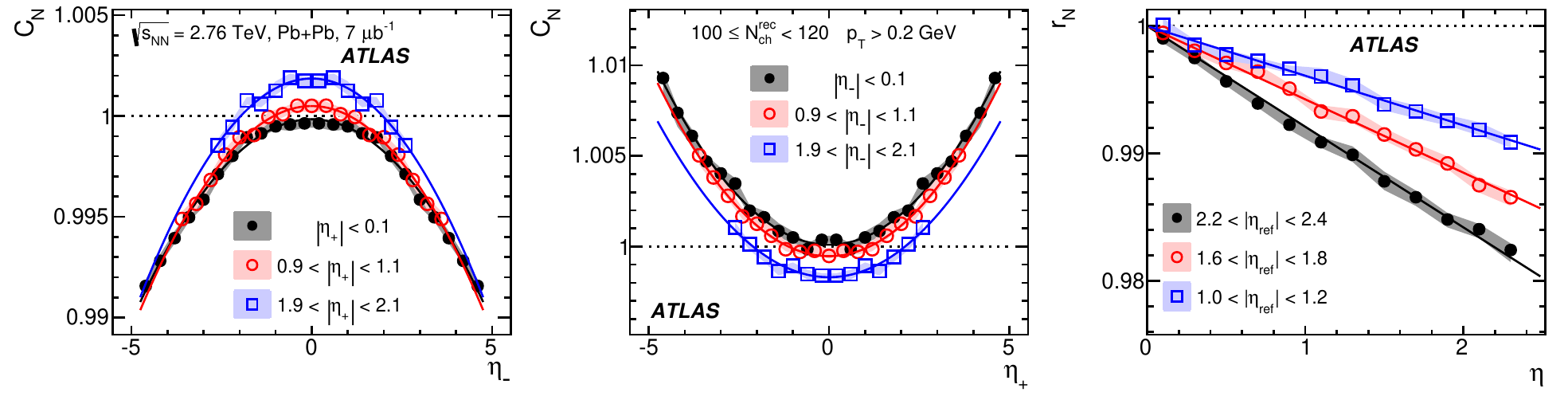}\vspace*{-0.1cm}
\includegraphics[width=1\linewidth]{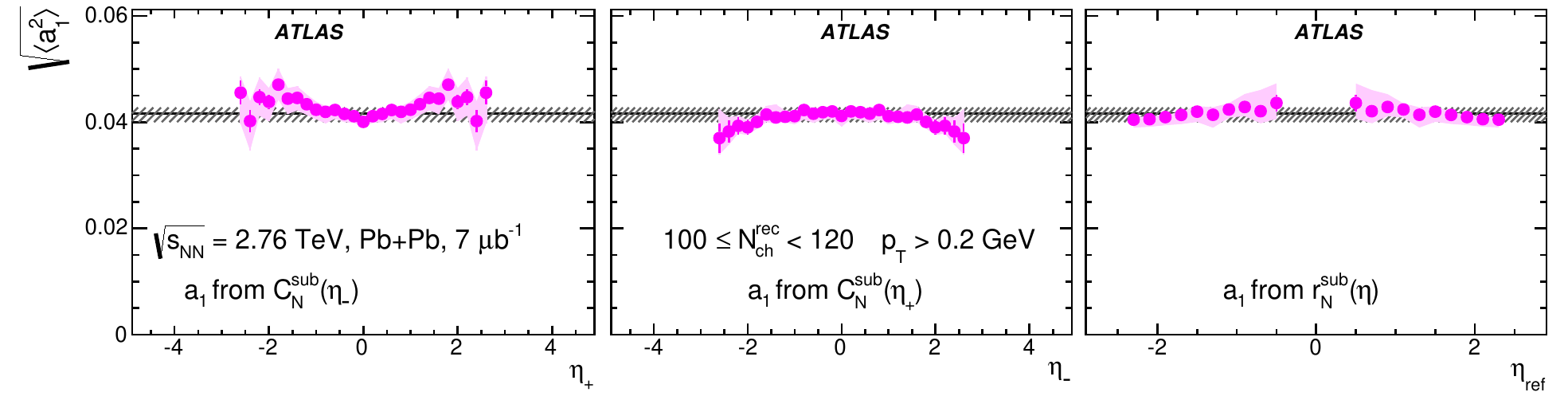}
\end{center}
\vspace*{-0.6cm}
\caption{\label{fig:7a} The distributions $C_{\rm N}^{\rm {sub}}(\eta_-)$ (top-left panel), $C_{\rm N}^{\rm {sub}}(\eta_+)$ (top-middle panel), and $r_{\rm N}^{\rm{sub}}(\eta)$ (top-right panel) obtained from $C_{\rm N}^{\rm {sub}}(\eta_1,\eta_2)$ in three ranges of $\eta_+$, $\eta_-$ and $\eta_{\mathrm{ref}}$, respectively, from Pb+Pb collisions with $100\leq\nchrec<120$. The solid lines indicate fits to either a quadratic function (top-left two panels) or a linear function (top-right panel). The $\lr{a_1^2}^{\nicefrac{1}{2}}$ values from the fits are shown in the corresponding lower panels as a function of the $\eta_+$, $\eta_-$, and $\eta_{\mathrm{ref}}$, respectively. The error bars and shaded bands represent the statistical and systematic uncertainties, respectively. The solid horizontal line and hashed band indicate the value and uncertainty of $\lr{a_1^2}^{\nicefrac{1}{2}}$ obtained from a Legendre expansion of the $C_{\rm N}^{\rm {sub}}(\eta_1,\eta_2)$.}
\end{figure}
\vspace*{-0.3cm}
\begin{figure}[!h]
\begin{center}
\includegraphics[width=1\linewidth]{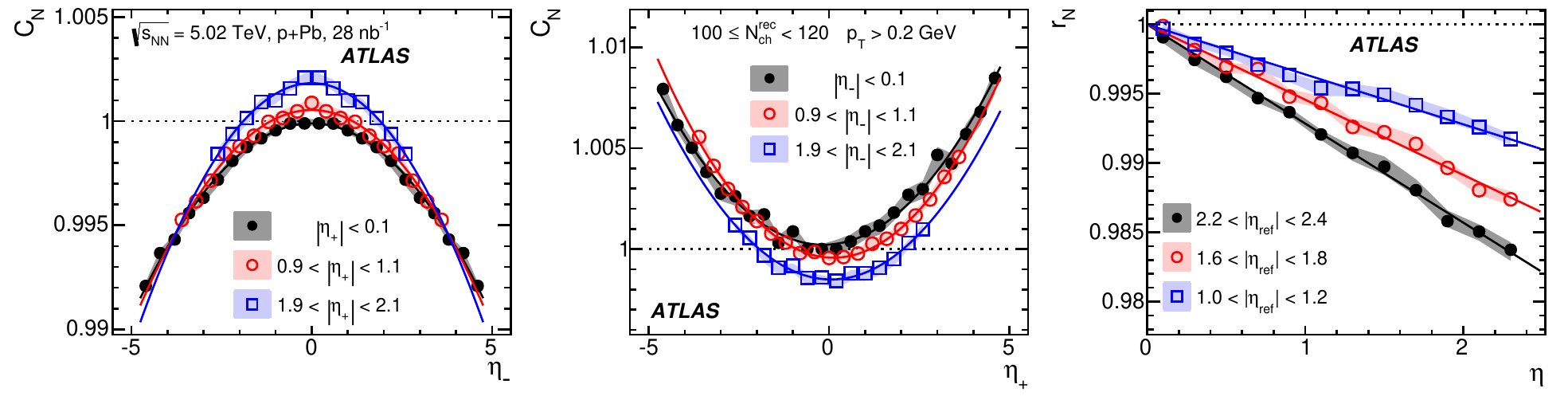}\vspace*{-0.1cm}
\includegraphics[width=1\linewidth]{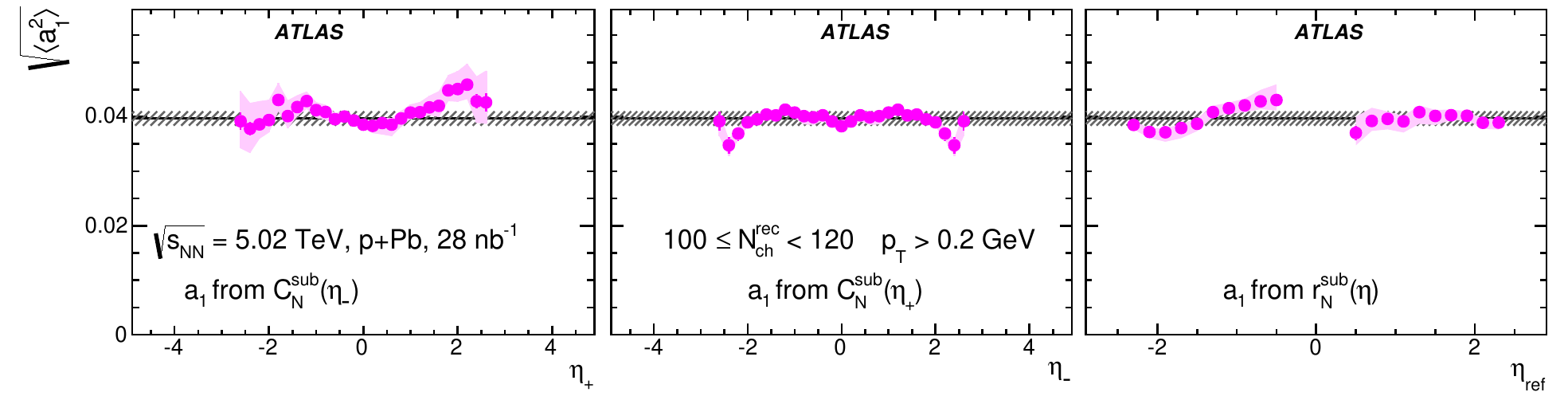}
\end{center}
\vspace*{-0.6cm}
\caption{\label{fig:7b} The distributions $C_{\rm N}^{\rm {sub}}(\eta_-)$ (top-left panel), $C_{\rm N}^{\rm {sub}}(\eta_+)$ (top-middle panel), and $r_{\rm N}^{\rm{sub}}(\eta)$ (top-right panel) obtained from $C_{\rm N}^{\rm {sub}}(\eta_1,\eta_2)$ in three ranges of $\eta_+$, $\eta_-$, and $\eta_{\mathrm{ref}}$, respectively, from $\pPb$ collisions with $100\leq\nchrec<120$. The solid lines indicate fits to either a quadratic function (top-left two panels) or a linear function (top-right panel). The $\lr{a_1^2}^{\nicefrac{1}{2}}$ values from the fits are shown in the corresponding lower panels as a function of the $\eta_+$, $\eta_-$, and $\eta_{\mathrm{ref}}$, respectively. The error bars and shaded bands represent the statistical and systematic uncertainties, respectively. The solid horizontal line and hashed band indicate the value and uncertainty of $\lr{a_1^2}^{\nicefrac{1}{2}}$ obtained from a Legendre expansion of the $C_{\rm N}^{\rm {sub}}(\eta_1,\eta_2)$.}
\end{figure}

\begin{figure}[!t]
\begin{center}
\includegraphics[width=1\linewidth]{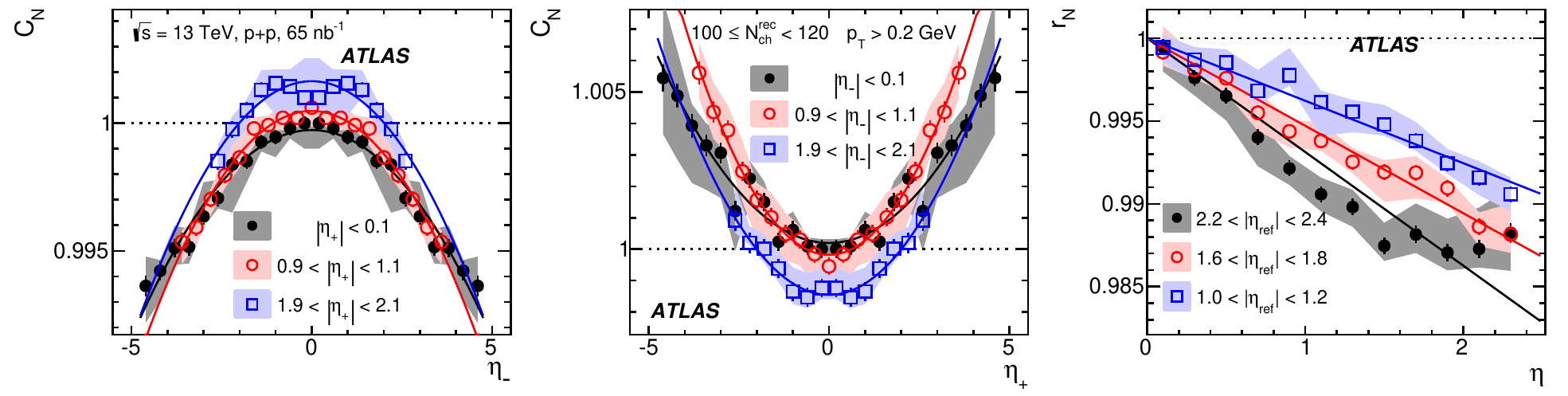}
\includegraphics[width=1\linewidth]{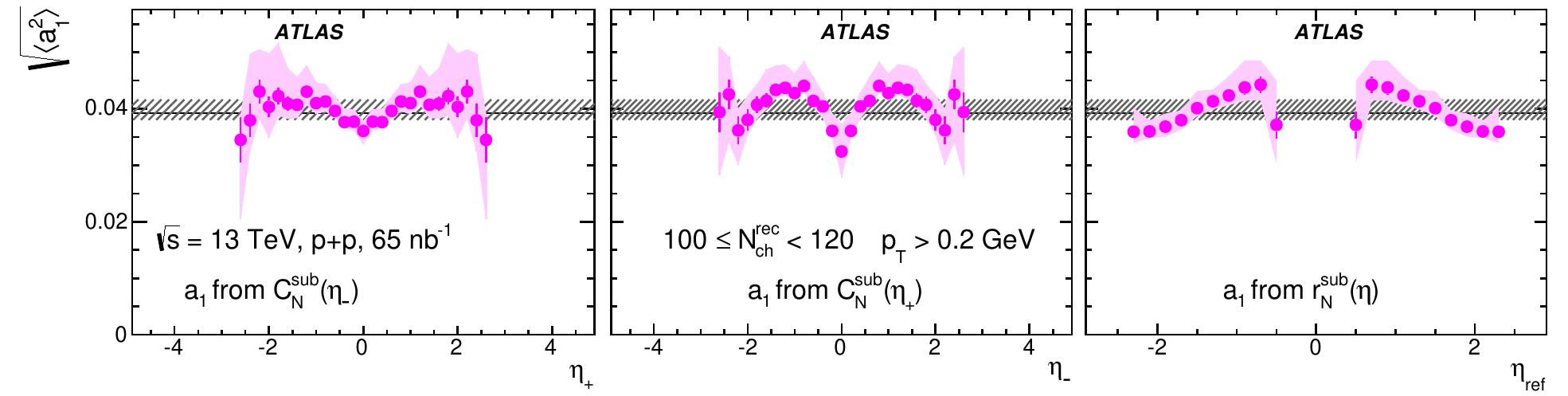}
\end{center}
\caption{\label{fig:7c} The distributions $C_{\rm N}^{\rm {sub}}(\eta_-)$ (top-left panel), $C_{\rm N}^{\rm {sub}}(\eta_+)$ (top-middle panel), and $r_{\rm N}^{\rm{sub}}(\eta)$ (top-right panel) obtained from $C_{\rm N}^{\rm {sub}}(\eta_1,\eta_2)$ in three ranges of $\eta_+$, $\eta_-$ and $\eta_{\mathrm{ref}}$, respectively, from $\pp$ collisions with $100\leq\nchrec<120$. The solid lines indicate fits to either a quadratic function (top-left two panels) or a linear function (top-right panel). The $\lr{a_1^2}^{\nicefrac{1}{2}}$ values from the fits are shown in the corresponding lower panels as a function of the $\eta_+$, $\eta_-$, and $\eta_{\mathrm{ref}}$, respectively. The error bars and shaded bands represent the statistical and systematic uncertainties, respectively. The solid horizontal line and hashed band indicate the value and uncertainty of $\lr{a_1^2}^{\nicefrac{1}{2}}$ obtained from a Legendre expansion of the $C_{\rm N}^{\rm {sub}}(\eta_1,\eta_2)$.}
\end{figure}

To quantify further the shape of the LRC in $C_{\rm N}^{\rm {sub}}(\eta_1,\eta_2)$, the $\lr{a_1^2}^{\nicefrac{1}{2}}$ coefficients are also calculated by fitting the 1-D distributions from the three projection methods as outlined in Sec.~\ref{sec:a1}: 1) quadratic fit of $C_{\rm N}^{\rm{sub}}(\eta_-)$ in a narrow range of $\eta_+$, 2) quadratic fit of $C_{\rm N}^{\rm{sub}}(\eta_+)$  in a narrow range of $\eta_-$, and 3) linear fit of $r_{\rm N}^{\rm{sub}}(\eta)$ in a narrow range of $\eta_{\rm{ref}}$. The results for Pb+Pb collisions with $100\leq\nchrec<120$ are shown in the first row of Fig.~\ref{fig:7a} for several selected projections and associated fits. The extracted $\lr{a_1^2}^{\nicefrac{1}{2}}$ values are shown in the bottom row as a function of the range of the projections. They are compared with the $\lr{a_1^2}^{\nicefrac{1}{2}}$ values obtained directly via the Legendre expansion of the entire $C_{\rm N}^{\rm {sub}}$ distribution, shown by the horizontal solid line. The $\lr{a_1^2}^{\nicefrac{1}{2}}$ values from all four methods are very similar. Figures~\ref{fig:7b} and \ref{fig:7c} show the same observables in $\pPb$ collisions and $\pp$ collisions, respectively. Results are quite similar to those in Pb+Pb collisions, albeit with larger systematic uncertainties arising from the subtraction of a larger short-range component. For $\pPb$ (Fig.~\ref{fig:7b}), the small FB asymmetry in the $C_{\rm N}^{\rm{sub}}$ distribution along the $\eta_+$ direction is responsible for the difference in $\lr{a_1^2}^{\nicefrac{1}{2}}$ between $\eta_+$ and $-\eta_+$ in the bottom-left panel and between $\eta_{\rm{ref}}$ and $-\eta_{\rm{ref}}$ in the bottom-right panel, but they still agree within their respective systematic uncertainties. 

\begin{figure}[!h]
\begin{center}
\includegraphics[width=1\linewidth]{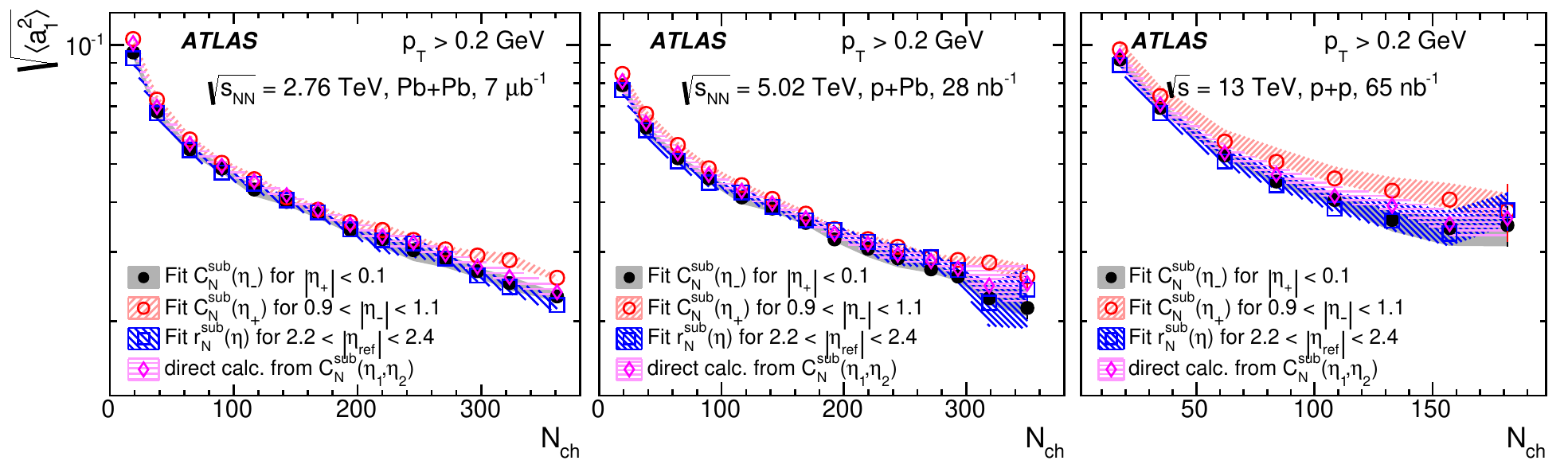}
\end{center}
\caption{\label{fig:9} The $\lr{a_1^2}^{\nicefrac{1}{2}}$ as a function of $\nch$ from four different methods, fit $C_{\rm N}^{\rm {sub}}(\eta_-)$ (solid circles), fit $C_{\rm N}^{\rm {sub}}(\eta_+)$ (open circles), fit $r_{\rm N}^{\rm{sub}}(\eta)$ (open squares), and Legendre expansion of $C_{\rm N}^{\rm {sub}}(\eta_1,\eta_2)$ (open diamonds), in Pb+Pb (left panel), $\pPb$ (middle panel), and $\pp$ (right panel) collisions. The error bars and shaded bands represent the statistical and systematic uncertainties, respectively.}
\end{figure}

Figure~\ref{fig:9} shows a comparison of the $\lr{a_1^2}^{\nicefrac{1}{2}}$ values extracted by the four methods as a function of $\nch$ in the three collision systems. Good agreement between the different methods is observed.

\begin{figure}[!t]
\begin{center}
\includegraphics[width=1\linewidth]{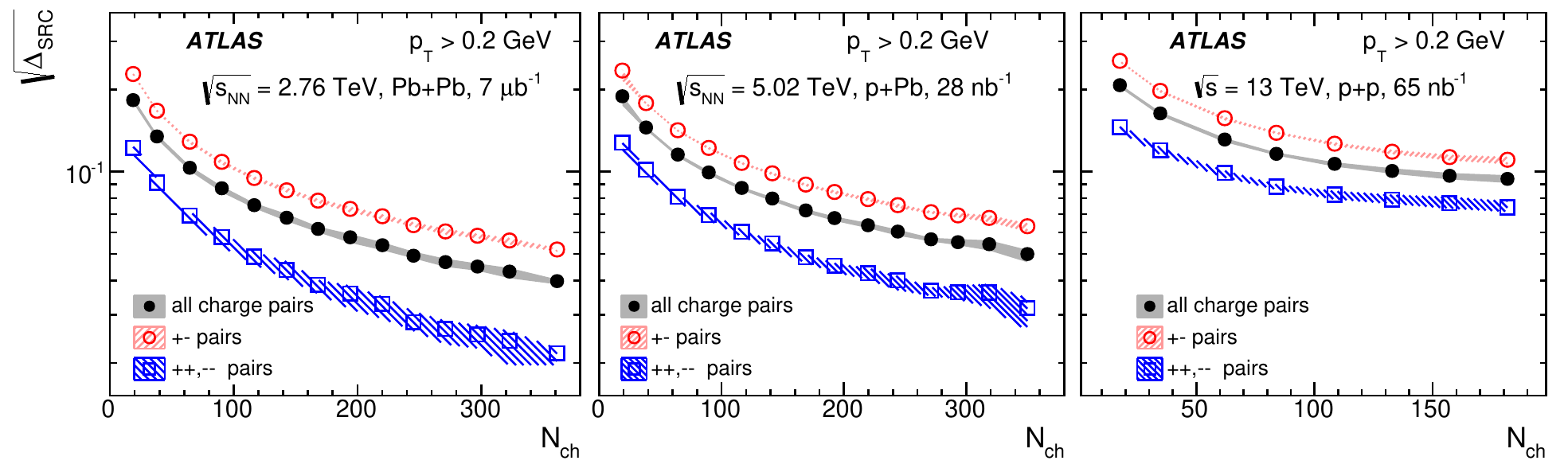}
\end{center}
\caption{\label{fig:11} The estimated magnitude of the short-range component $\sqrt{\Delta_{\rm {SRC}}}$ as a function of $\nch$ for all-charge (solid circles), opposite-charge (open circles), and same-charge (open squares) pairs in Pb+Pb (left panel), $\pPb$ (middle panel), and $\pp$ (right panel) collisions. The shaded bands represent the systematic uncertainties, and the statistical uncertainties are smaller than the symbols.}
\end{figure}

On the other hand, the SRC is expected to have strong dependence on the charge combinations and collision systems, as shown by Figs.~\ref{fig:5} and \ref{fig:6}. The magnitude of the SRC is quantified by $\delta_{\mathrm{SRC}}(\eta_1,\eta_2)$ averaged over the two-particle pseudorapidity phase space:
\begin{eqnarray}\label{eq:e0}
\Delta_{\mathrm{SRC}} = \frac{\int_{-Y}^{Y} \delta_{\rm {SRC}}(\eta_1,\eta_2) \,d\eta_1\,d\eta_2}{4Y^2}\;.
\end{eqnarray}
The corresponding contribution of the SRC at the single-particle level is $\sqrt{\Delta_{\mathrm{SRC}}}$, which can be directly compared with the strength of the LRC characterized by $\lr{a_1^2}^{\nicefrac{1}{2}}$. Figure~\ref{fig:11} shows the values of $\sqrt{\Delta_{\mathrm{SRC}}}$ as a function of $\nch$ for different charge combinations in the three collision systems. The strength of the SRC always decreases with $\nch$, and it is larger for smaller collision systems and opposite-charge pairs.

\begin{figure}[!t]
\begin{center}
\includegraphics[width=1\linewidth]{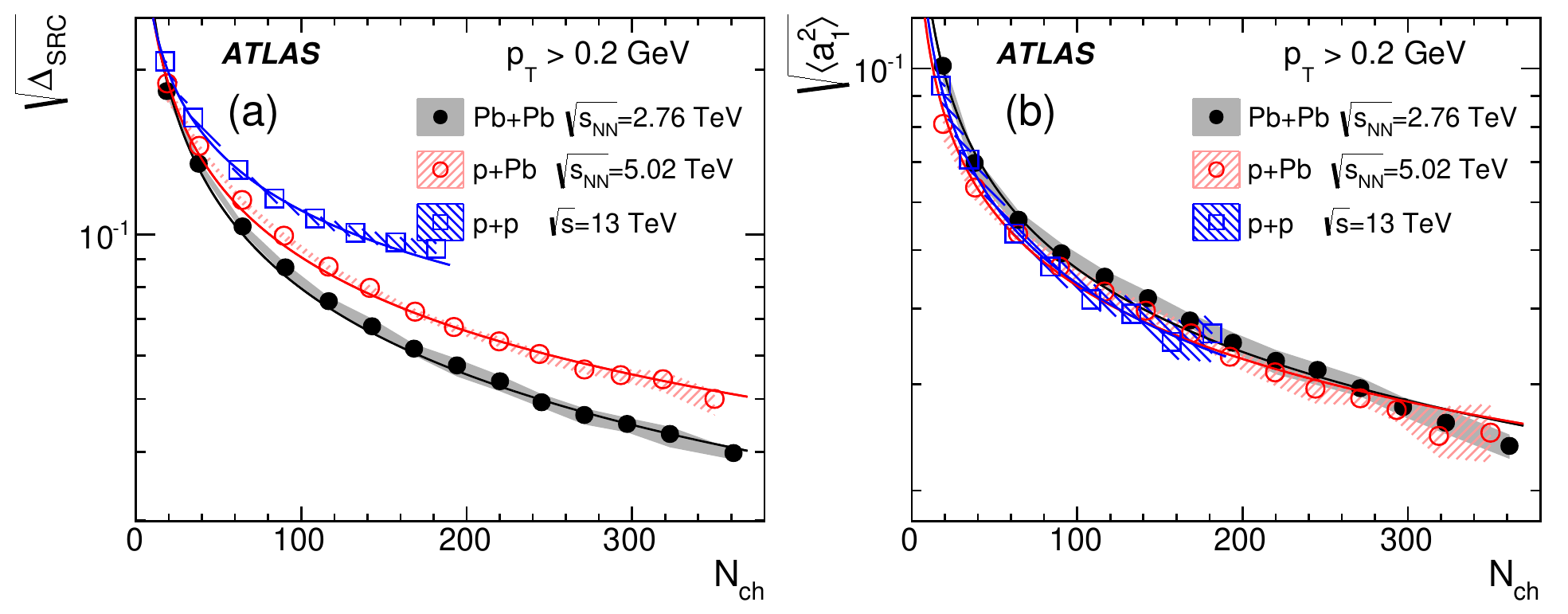}
\end{center}
\caption{\label{fig:12} The estimated magnitude of the short-range component $\sqrt{\Delta_{\rm {SRC}}}$ (left panel) and $\lr{a_1^2}^{\nicefrac{1}{2}}$ (right panel) values as a function of $\nch$ for all-charge pairs in Pb+Pb (solid circles), $\pPb$ (open circles), and $\pp$ (open squares) collisions. The shaded bands represent the systematic uncertainties, and the statistical uncertainties are smaller than the symbols.}
\end{figure}

Figure~\ref{fig:12} compares the strength of the SRC in terms of $\sqrt{\Delta_{\rm {SRC}}}$ and the LRC in terms of $\lr{a_1^2}^{\nicefrac{1}{2}}$ for the three collision systems. The values of $\sqrt{\Delta_{\rm {SRC}}}$ are observed to differ significantly while the values of $\lr{a_1^2}^{\nicefrac{1}{2}}$ agree within $\pm10\%$ between the three collision systems. 

The strength of the SRC and LRC can be related to the number of clusters $n$ contributing to the final multiplicity $\nch$, where $n$ is the sum of clusters from the projectile and target nucleon or nucleus, $n=n_{\rm{F}}+n_{\rm{B}}$. The LRC is expected to be related to the asymmetry between $n_{\rm{F}}$ and $n_{\rm{B}}$:
\begin{eqnarray}\label{eq:e3}
A_n = \frac{n_{\rm{F}}-n_{\rm{B}}}{n_{\rm{F}}+n_{\rm{B}}},\;\;\lr{a_1^2} \propto \lr{A_n^2}\;.
\end{eqnarray}
The clusters could include the participating nucleons, subnucleonic degrees of freedom such as the fragmentation of scattered partons, or resonance decays. In an independent cluster model~\cite{Berger:1974vn}, each cluster emits the same number of pairs and the number of clusters follows Poisson fluctuations. In this picture, both the SRC in terms of $\Delta_{\rm {SRC}}$ and LRC in terms of $\lr{a_1^2}$ should scale approximately as the inverse of the number of clusters, and hence, assuming $n$ and $\nch$ are proportional, the $\sqrt{\Delta_{\rm {SRC}}}$ and $\lr{a_1^2}^{\nicefrac{1}{2}}$ values in Fig.~\ref{fig:12} are expected to follow a simple power-law function in $\nch$:
\begin{eqnarray}\label{eq:e4}
\sqrt{\Delta_{\rm {SRC}}} \sim \lr{a_1^2}^{\nicefrac{1}{2}} \sim \frac{1}{n^{\alpha}}\sim \frac{1}{\nch^{\alpha}}\;, \alpha\approx 0.5\;.
\end{eqnarray}
A power index that is less than one half, $\alpha<0.5$, would suggest that $n$ grows more slowly than $\nchrec$, and vice versa.

To test this idea, the $\sqrt{\Delta_{\rm {SRC}}}$ and $\lr{a_1^2}^{\nicefrac{1}{2}}$ data in Fig.~\ref{fig:12} are fit to a power-law function: $c/\nch^{\alpha}$. The function describes the $\nch$ dependence in all three collision systems, with a reduced $\chi^2$ values ranging between 0.2 and 0.9. The extracted power index values are summarized in Table~\ref{tab:fit}. The values of $\alpha$ for the SRC are found to be smaller for smaller collision systems, they are close to 0.5 in the Pb+Pb collisions and are significantly smaller than 0.5 in the $\pp$ collisions. In contrast, the values of $\alpha$ for $\lr{a_1^2}^{\nicefrac{1}{2}}$ agree within uncertainties between the three systems and are slightly below 0.5. 

\begin{table}[!h]
\caption{\label{tab:fit} The power index and associated total uncertainty from a power-law fit of the $\nch$ dependence of $\sqrt{\Delta_{\rm {SRC}}}$ and $\lr{a_1^2}^{\nicefrac{1}{2}}$.}
\centering
\begin{tabular}{|c|c|c|c|}\hline
                                        & Pb+Pb            & $\pPb$         & $\pp$            \tabularnewline\hline
$\alpha$ for $\sqrt{\Delta_{\rm {SRC}}}$  & $0.505\pm0.011$ & $0.450\pm0.010$ & $0.365\pm0.014$ \tabularnewline\hline
$\alpha$ for $\lr{a_1^2}^{\nicefrac{1}{2}}$         & $0.454\pm0.011$ & $0.433\pm0.014$ & $0.465\pm0.018$ \tabularnewline\hline
\end{tabular}
\end{table}

\begin{figure}[!t]
\begin{center}
\includegraphics[width=1\linewidth]{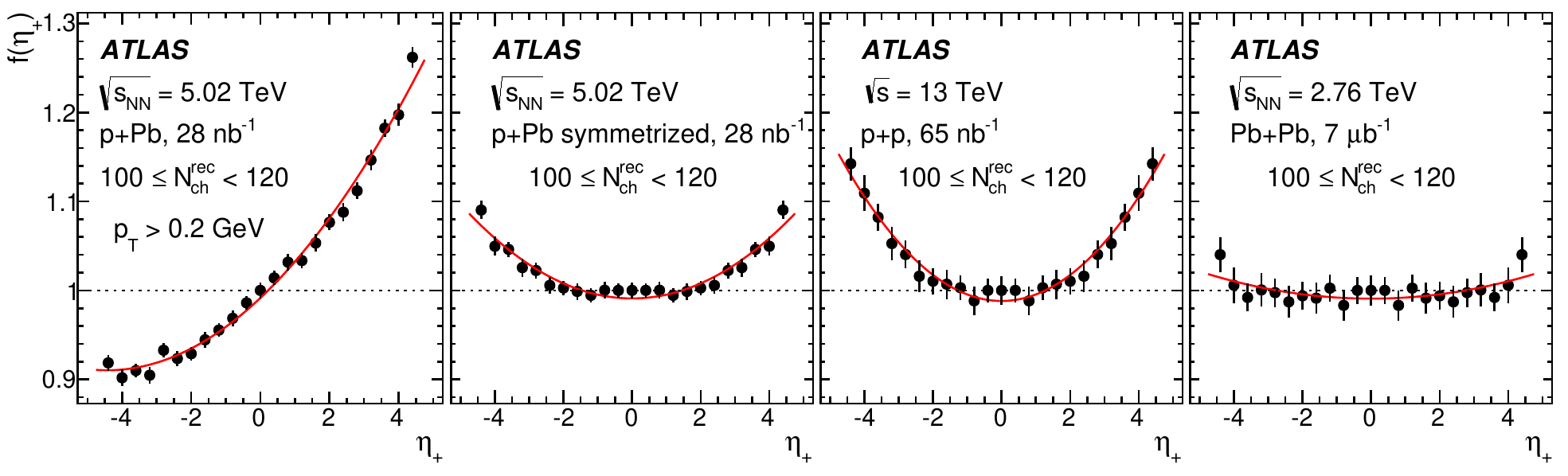}
\end{center}
\caption{\label{fig:13} The shape of the SRC in $\eta_+$ represented by $f(\eta_+)$ calculated via Eq.~\eqref{eq:b1} for $\pPb$, symmetrized-$\pPb$, $\pp$, and Pb+Pb collisions with $100\leq\nchrec<120$. The solid lines represent a fit to a quadratic function.}
\end{figure}

One striking feature of the correlation function in $\pPb$ collisions, for example in Fig.~\ref{fig:5}, is a large FB asymmetry of the SRC, $\delta_{\rm {SRC}}(\eta_1,\eta_2)$ along the $\eta_{+}$ direction. Even in $\pp$ collisions, the $\delta_{\rm {SRC}}$ distribution is not uniform, but instead shows a quadratic increase towards large $|\eta_{+}|$ values. According to the discussion in Sec.~\ref{sec:corr}, the shape of the $\delta_{\rm {SRC}}$ distribution in $\eta_+$ is described by the $f(\eta_+)$ defined in Eq.~\eqref{eq:b1}. Examples of the $f(\eta_+)$ are shown in Fig.~\ref{fig:13} for $\pPb$, symmetrized-$\pPb$,  $\pp$, and Pb+Pb collisions with $100~\leq~\nchrec~<~120$. As described in Sec.~\ref{sec:corr}, symmetrized-$\pPb$ results are obtained by averaging the proton-going and lead-going directions such that $C(\eta_1,\eta_2)=C(-\eta_1,-\eta_2)$.

The independent cluster picture discussed above offers a simple interpretation of the shape of $f(\eta_+)$. Assuming the population of clusters is a function of $\eta$, $n_{\rm{c}}(\eta)$, and on average each cluster produces $m$ charged particles according to a Poisson distribution, then the number of the SRC pairs scales as $n_{\rm{c}} \lr{m(m-1)} = n_{\rm{c}} \lr{m}^2$ and the number of the combinatorial pairs scales as $\left(\vphantom{A^{b}}n_{\rm{c}}\lr{m}\right)^2$. Therefore the strength of the SRC at given $\eta$ is expected to scale as:
\begin{eqnarray}
\label{eq:e5}
\delta_{\rm {SRC}}(\eta,\eta) \propto \frac{n_{\rm{c}} \lr{m(m-1)} }{\left(\vphantom{A^{b}}n_{\rm{c}}\lr{m}\right)^2} = \frac{1}{n_{\rm{c}}} \propto \frac{1}{\,d\nch/\,d\eta}
\end{eqnarray}
where $n_{\rm{c}}(\eta)$ is assumed to be proportional to the local charge-particle multiplicity density $\,d\nch/\,d\eta$. Hence the fact that $f(\eta_+)$ is larger in the proton-going direction than in the Pb-going direction in $\pPb$ collisions simply reflects the asymmetric shape of the $\,d\nch/\,d\eta$ distribution in each event~\cite{Aad:2015zza}. The quadratic shape of $f(\eta_+)$ for $\pp$ and symmetrized-$\pPb$ system therefore reflects a large, intrinsic FB asymmetry of $d\nch/d\eta$ on an event-by-event level. The FB asymmetry in $\pp$ collisions is slightly larger than $\pPb$ collisions at comparable $\nch$, but is significantly less in Pb+Pb collisions. This observation suggests that the FB asymmetry for particle production in $\pp$ collisions could be as large as that in $\pPb$ collisions at comparable event activity, whereas the FB asymmetry for particle production is smaller in Pb+Pb collisions.

\section{Comparison to models}

QCD-inspired models such as PYTHIA and EPOS are often used to describe the particle production in $\pp$ collisions. ATLAS has previously compared the predictions of the PTYHIA8 A2 and EPOS LHC tunes with various single particle distributions, such as the $\pT$, $\eta$ and the event-by-event $\nch$ distributions, fully unfolded for detector effects~\cite{Aad:2016mok,Aaboud:2016itf}. Reasonable agreement has been observed for these single-particle observables.  In order to perform a data model comparison, the multiplicity correlation procedure used on the data is repeated for the two models to extract the SRC and LRC components. The extracted LRC in these models is then decomposed into Legendre coefficients of different order. The coefficients are found to be dominated by $\lr{a_1^2}^{\nicefrac{1}{2}}$, consistent with the observation that the shapes of the LRC are similar to those in the $\pp$ data in Fig.~\ref{fig:5}. However the values of $\lr{a_1^2}^{\nicefrac{1}{2}}$ predicted by the models are found to be much smaller than the $\pp$ data at the same $\nch$.

\begin{figure}[!h]
\begin{center}
\includegraphics[width=1\linewidth]{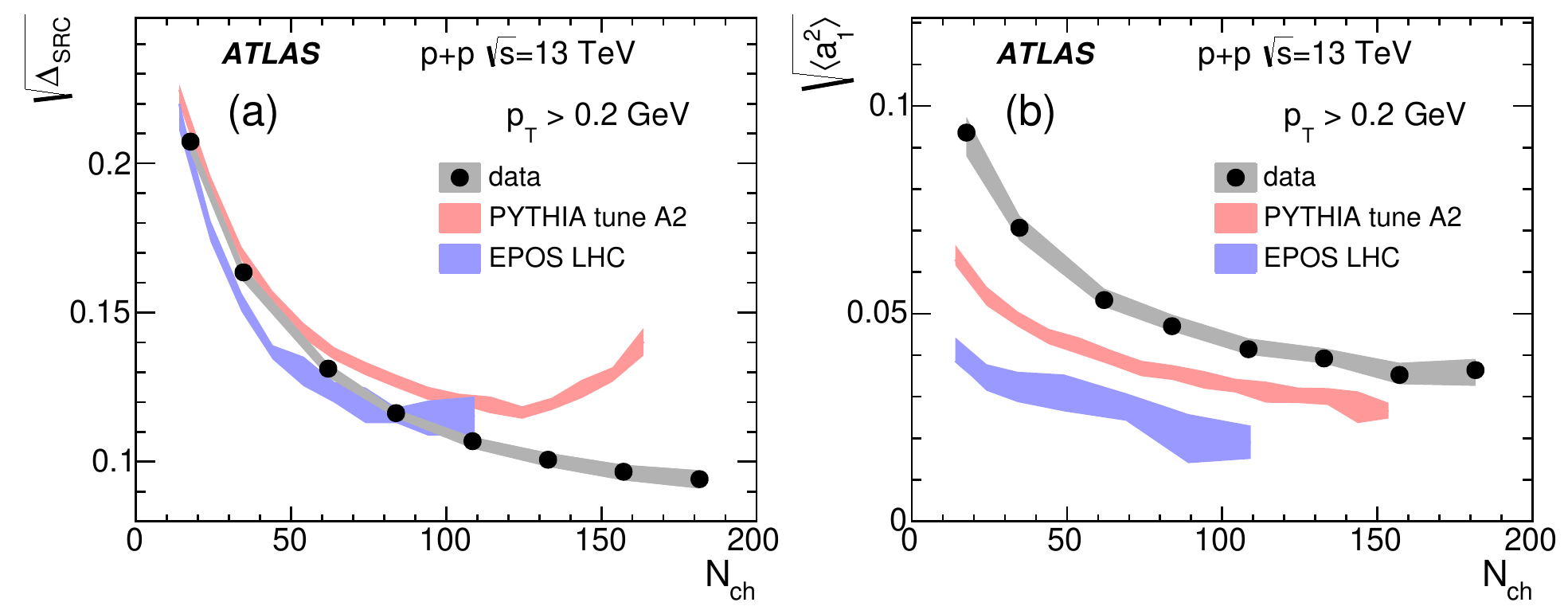}
\end{center}
\caption{\label{fig:14} The $\sqrt{\Delta_{\rm {SRC}}}$ (left panel) and $\lr{a_1^2}^{\nicefrac{1}{2}}$ (right panel) as a function of $\nch$ in $\pp$ collisions at $\sqrt{s}=13$ TeV, compared between data and PYTHIA 8 A2 and EPOS LHC. The shaded bands represent the total uncertainties.}
\end{figure}

For a more direct comparison, Figure~\ref{fig:14} show the $\nch$ dependence of SRC and LRC from the data and the two models in $\pp$ collisions. The systematic uncertainties on the model predictions are dominated by the uncertainty in separating the SRC and LRC, as discussed in Sec.~\ref{sec:syscheck}. However at large $\nch$, they are also limited by the available MC statistics. There is some indication that the values of $\sqrt{\Delta_{\rm {SRC}}}$ from data are larger than the EPOS predictions and smaller than those from PYTHIA 8. Furthermore, the values from PYTHIA 8 increase for $\nch>120$, a trend not supported by the data. On the other hand, both models underestimate significantly the values of $\lr{a_1^2}^{\nicefrac{1}{2}}$, suggesting that the FB multiplicity fluctuations in both models are significantly weaker than in the $\pp$ data. Therefore these two models, which were tuned to describe many single particle observables, fail to describe the longitudinal correlations between the produced charged particles.

\section{Summary}
Two-particle pseudorapidity correlations are measured with the ATLAS detector in $\sqrtsnn$ = 2.76 $\TeV$ Pb+Pb, $\sqrtsnn$ = 5.02 $\TeV$ $\pPb$, and $\sqrt{s}$ = 13~$\TeV$ $\pp$ collisions at the LHC, with total integrated luminosities of approximately 7~$\mu\mathrm{b}^{-1}$, 28~$\mathrm{nb}^{-1}$, and 65~$\mathrm{nb}^{-1}$, respectively. The correlation function $C_{\rm N}(\eta_1,\eta_2)$ is measured using charged particles in the pseudorapidity range $|\eta|<2.4$ with transverse momentum $\pT>0.2$~$\GeV$, and it is measured as a function of event multiplicity $\nch$ defined by the total number of charged particles with $|\eta|<2.5$ and $\pT>0.4$~$\GeV$. The correlation function shows an enhancement along the $\eta_1\approx\eta_2$ direction and suppression at $\eta_1\approx -\eta_2\sim\pm2.4$, consistent with the expectation from an event-by-event forward-backward asymmetry in the multiplicity fluctuation (the long-range correlations or LRC). However, the correlation function also has a large narrow ``ridge'' along the $\eta_1\approx\eta_2$ direction associated with short-range correlations (SRC). The magnitudes of the SRC in $\pPb$ is found to be larger in the proton-going direction than the lead-going direction, reflecting the fact that the particle multiplicity is smaller in the proton-going direction. This is consistent with the observation that the SRC strength increases for smaller $\nch$. The SRC is observed to be much stronger for opposite-charge pairs than for the same-charge pairs, while the LRC is found to be similar for the two charge combinations. Based on this, a data-driven subtraction method was developed to separate the SRC and the LRC. The magnitudes of the SRC and the LRC are then compared for the three collision systems at similar values of $\nch$.

After subtracting out the SRC $\delta_{\mathrm{SRC}}(\eta_1,\eta_2)$, the correlation function $C_{\rm N}^{\rm {sub}}(\eta_1,\eta_2)$ is decomposed into a sum of products of Legendre polynomials that describe the different shape components, and the coefficients $\lr{a_na_m}$ are calculated. Significant values are observed for $\lr{a_1^2}$ in all $\nch$ ranges and higher-order coefficients are consistent with zero, and suggesting that $C_{\rm N}^{\rm {sub}}$ has an approximate functional form $C_{\rm N}^{\rm {sub}}\approx 1+\lr{a_1^2}\eta_1\eta_2$.  The quantity $\lr{a_1^2}$ is also estimated by parameterization of the shape of the correlation function in narrow ranges of $\eta_-=\eta_1-\eta_2$ and $\eta_+=\eta_1+\eta_2$, or from a ratio $C_{\rm N}^{\rm {sub}}(\eta_1,\eta_2)/C_{\rm N}^{\rm {sub}}(-\eta_1,\eta_2)$, and consistent results are obtained. The magnitude of the SRC and $\lr{a_1^2}^{\nicefrac{1}{2}}$ are compared for the three collision systems as a function of $\nch$. Large differences are observed for the SRC, but the values of $\lr{a_1^2}^{\nicefrac{1}{2}}$ agree within $\pm10$\% at the same $\nch$. The $\nch$ dependences of both the SRC and $\lr{a_1^2}^{\nicefrac{1}{2}}$ follow an approximate power-law shape. The power index for $\lr{a_1^2}^{\nicefrac{1}{2}}$ is approximately the same for the three collision systems. In contrast, the power-law index for the SRC is smaller for smaller collision systems. The SRC distribution shows strong dependence on $\eta_+$ in $\pPb$ and $\pp$, but much weaker dependence in Pb+Pb collisions. The $\delta_{\mathrm{SRC}}(\eta_+)$ distribution, after symmetrizing the proton and lead directions, is found to be similar to the SRC in $\pp$ collisions with comparable $\nch$, suggesting that the event-by-event FB asymmetry for particle production is similar in $\pp$ and $\pPb$ collisions with comparable event activity. The PYTHIA 8 A2 and EPOS LHC models, which were tuned to describe many single particle observables in $\pp$ collisions, fail to describe the SRC and the LRC observed in the $\pp$ data.
\section*{Acknowledgements}

We thank CERN for the very successful operation of the LHC, as well as the
support staff from our institutions without whom ATLAS could not be
operated efficiently.

We acknowledge the support of ANPCyT, Argentina; YerPhI, Armenia; ARC, Australia; BMWFW and FWF, Austria; ANAS, Azerbaijan; SSTC, Belarus; CNPq and FAPESP, Brazil; NSERC, NRC and CFI, Canada; CERN; CONICYT, Chile; CAS, MOST and NSFC, China; COLCIENCIAS, Colombia; MSMT CR, MPO CR and VSC CR, Czech Republic; DNRF and DNSRC, Denmark; IN2P3-CNRS, CEA-DSM/IRFU, France; GNSF, Georgia; BMBF, HGF, and MPG, Germany; GSRT, Greece; RGC, Hong Kong SAR, China; ISF, I-CORE and Benoziyo Center, Israel; INFN, Italy; MEXT and JSPS, Japan; CNRST, Morocco; FOM and NWO, Netherlands; RCN, Norway; MNiSW and NCN, Poland; FCT, Portugal; MNE/IFA, Romania; MES of Russia and NRC KI, Russian Federation; JINR; MESTD, Serbia; MSSR, Slovakia; ARRS and MIZ\v{S}, Slovenia; DST/NRF, South Africa; MINECO, Spain; SRC and Wallenberg Foundation, Sweden; SERI, SNSF and Cantons of Bern and Geneva, Switzerland; MOST, Taiwan; TAEK, Turkey; STFC, United Kingdom; DOE and NSF, United States of America. In addition, individual groups and members have received support from BCKDF, the Canada Council, CANARIE, CRC, Compute Canada, FQRNT, and the Ontario Innovation Trust, Canada; EPLANET, ERC, FP7, Horizon 2020 and Marie Sk{\l}odowska-Curie Actions, European Union; Investissements d'Avenir Labex and Idex, ANR, R{\'e}gion Auvergne and Fondation Partager le Savoir, France; DFG and AvH Foundation, Germany; Herakleitos, Thales and Aristeia programmes co-financed by EU-ESF and the Greek NSRF; BSF, GIF and Minerva, Israel; BRF, Norway; Generalitat de Catalunya, Generalitat Valenciana, Spain; the Royal Society and Leverhulme Trust, United Kingdom.

The crucial computing support from all WLCG partners is acknowledged gratefully, in particular from CERN, the ATLAS Tier-1 facilities at TRIUMF (Canada), NDGF (Denmark, Norway, Sweden), CC-IN2P3 (France), KIT/GridKA (Germany), INFN-CNAF (Italy), NL-T1 (Netherlands), PIC (Spain), ASGC (Taiwan), RAL (UK) and BNL (USA), the Tier-2 facilities worldwide and large non-WLCG resource providers. Major contributors of computing resources are listed in Ref.~\cite{ATL-GEN-PUB-2016-002}.

\clearpage 

\printbibliography
\newpage 
\begin{flushleft}
{\Large The ATLAS Collaboration}

\bigskip

M.~Aaboud$^{\rm 135d}$,
G.~Aad$^{\rm 86}$,
B.~Abbott$^{\rm 113}$,
J.~Abdallah$^{\rm 64}$,
O.~Abdinov$^{\rm 12}$,
B.~Abeloos$^{\rm 117}$,
R.~Aben$^{\rm 107}$,
O.S.~AbouZeid$^{\rm 137}$,
N.L.~Abraham$^{\rm 149}$,
H.~Abramowicz$^{\rm 153}$
% ...
\bigskip
\\
% ...
$^{12}$ Institute of Physics, Azerbaijan Academy of Sciences, Baku, Azerbaijan\\
% ...
$^{64}$ University of Iowa, Iowa City IA, United States of America\\
% ...
$^{86}$ CPPM, Aix-Marseille Universit{\'e} and CNRS/IN2P3, Marseille, France\\
% ...
$^{107}$ Nikhef National Institute for Subatomic Physics and University of Amsterdam, Amsterdam, Netherlands\\
% ...
$^{113}$ Homer L. Dodge Department of Physics and Astronomy, University of Oklahoma, Norman OK, United States of America\\
% ...
$^{117}$ LAL, Univ. Paris-Sud, CNRS/IN2P3, Universit{\'e} Paris-Saclay, Orsay, France\\
% ...
$^{135}$ $^{(a)}$ Facult{\'e} des Sciences Ain Chock, R{\'e}seau Universitaire de Physique des Hautes Energies - Universit{\'e} Hassan II, Casablanca; $^{(b)}$ Centre National de l'Energie des Sciences Techniques Nucleaires, Rabat; $^{(c)}$ Facult{\'e} des Sciences Semlalia, Universit{\'e} Cadi Ayyad, LPHEA-Marrakech; $^{(d)}$ Facult{\'e} des Sciences, Universit{\'e} Mohamed Premier and LPTPM, Oujda; $^{(e)}$ Facult{\'e} des sciences, Universit{\'e} Mohammed V, Rabat, Morocco\\
% ...
$^{137}$ Santa Cruz Institute for Particle Physics, University of California Santa Cruz, Santa Cruz CA, United States of America\\
% ...
$^{149}$ Department of Physics and Astronomy, University of Sussex, Brighton, United Kingdom\\
% ...
$^{153}$ Raymond and Beverly Sackler School of Physics and Astronomy, Tel Aviv University, Tel Aviv, Israel\\
% ...
$^{a}$ Also at Department of Physics, King's College London, London, United Kingdom\\
$^{b}$ Also at Institute of Physics, Azerbaijan Academy of Sciences, Baku, Azerbaijan\\
$^{c}$ Also at Novosibirsk State University, Novosibirsk, Russia\\
$^{d}$ Also at TRIUMF, Vancouver BC, Canada\\
$^{e}$ Also at Department of Physics {\&} Astronomy, University of Louisville, Louisville, KY, United States of America\\
% ...
\end{flushleft}

 
\end{document}